\documentclass[fleqn,usenatbib]{mnras}

\usepackage{newtxtext,newtxmath}

\usepackage[T1]{fontenc}

\DeclareRobustCommand{\VAN}[3]{#2}
\let\VANthebibliography\thebibliography
\def\thebibliography{\DeclareRobustCommand{\VAN}[3]{##3}\VANthebibliography}

\usepackage{enumitem}

\usepackage{graphicx} 
\usepackage{amsmath} 
\usepackage{subcaption}
\usepackage{placeins}
\usepackage[table]{xcolor}
\usepackage[nameinlink,noabbrev]{cleveref}
\creflabelformat{equation}{#2\textup{(#1)}#3}
\crefname{figure}{Fig.}{Figs.}
\crefname{table}{Table}{Tables}

\title[AGN energetics from remnant radio galaxies]{AGN energetics and lifetimes from remnant radio galaxies}

\author[B. Quici et al.]{Benjamin Quici$^{1}$, Ross J. Turner$^{2}$\thanks{Email:ross.turner@utas.edu.au}, Nicholas Seymour$^{1}$, Natasha Hurley-Walker$^{1}$\\
$^{1}$International Centre for Radio Astronomy Research, Curtin University, Bentley, WA 6102, Australia\\
$^{2}$School of Natural Sciences, University of Tasmania, Private Bag 37, Hobart, 7001, Australia}

\date{Accepted 2024 December 30. Received 2024 December 08; in original form 2024 May 20}

\pubyear{2024}

\begin{document}
\label{firstpage}
\pagerange{\pageref{firstpage}--\pageref{lastpage}}
\maketitle

\begin{abstract}
The energy coupling efficiency of active galactic nucleus (AGN) outbursts is known to {vary} significantly with factors including the jet kinetic power, duration of the outburst, and properties of the host galaxy {group or} cluster. As such, constraints on their jet power and lifetime functions are crucial to quantify the role of kinetic-mode AGN feedback on the evolution of galaxies since $z \sim 1$. In this work, we address this issue by measuring the energetics of a sample of 79~low-redshift (0.02 $< z <$ 0.2) remnant radio galaxies compiled from large-sky radio surveys -- {remnants} uniquely probe the full duration of an AGN outburst. The jet kinetic power and outburst duration of each remnant are determined using the RAiSE dynamical model based on the surface brightness distribution observed in multi-frequency radio images. We compare the energetics constrained for this sample to those predicted for mock radio source populations -- with various intrinsic functions for jet power and lifetime distributions -- to correct for sample selection biases imposed on our sample. The intrinsic jet power and lifetime functions that yield a selection-biased mock population most similar to our observed sample are determined using Bayesian inference. Our analysis places robust constraints on assumed power-law indices for the intrinsic jet power and lifetime functions: $p(Q)\propto Q^{-1.49\pm0.07}$ and $p(t_{\mathrm{on}})\propto t_{\mathrm{on}}^{-0.97\pm0.12}$ respectively. We discuss the implications of these findings for feedback-regulated accretion and the self-regulating nature of jet activity. The methodology proposed in this work can be extended to active radio galaxies in future studies.
\end{abstract}




\section{Introduction}
\label{sec:ch4_introduction}
In the concordance view of galaxy formation and evolution, relativistic jets associated with radio-loud active galactic nuclei (AGNs) play a critical role in regulating the gas residing in the interstellar and circumgalactic environments. On small (galactic) scales, the jets are capable of dispelling large amounts of molecular gas \citep[e.g.,][]{2008A&A...491..407N,2011A&A...533L..10D,2013Sci...341.1082M,2014A&A...572A..40E,2015A&A...580A...1M,2018A&A...618A...6K,2021A&A...656A..55M}, and can seemingly promote \citep{2008A&A...491..407N,2012A&A...541L...7D,2013Sci...341.1082M,2021AN....342.1140M} or suppress \citep{2006ApJ...647.1040C,2009MNRAS.396...61T,2012MNRAS.421.1603C,2017ApJ...844...37D} star formation. On larger scales, the buoyant rise of jet-inflated bubbles excavate cavities in their hot intracluster environments \citep[e.g.,][]{2002MNRAS.332..729C}, which act to suppress the cooling flows at the centre of cool-core clusters \citep{2003MNRAS.344L..43F,2005ApJ...635..894F,2007ARA&A..45..117M,2009A&A...501..835M,2012NewAR..56...93A,2012ARA&A..50..455F}. Recent findings also show that `maintenance-mode feedback', the self-regulating mechanism by which the jets maintain a heating/cooling balance in their hot atmospheres, must be invoked in galaxy formation models to explain, in particular, the suppression of star formation in the most massive galaxies \citep[e.g.,][]{2006MNRAS.370..645B,2006MNRAS.365...11C,2009ApJ...699..525S,2017MNRAS.471..658R}. 

Despite the various successes of modern galaxy formation models \citep[e.g.,][]{2014MNRAS.440.1590D,2014MNRAS.444.1518V,2015MNRAS.446..521S}, constraints from observations are still required to quantify the global energetic impact of AGNs, and to elucidate the connection between the environment and fuelling of the supermassive black hole (SMBH). One valuable insight lies in the observed mass-dependence of the radio loud fraction, $f_{\mathrm{RL}}$, which rises with the stellar mass of the host galaxy \citep{2009AN....330..184B} and likely achieves ubiquity ($\approx 100\%$) amongst the most massive galaxies \citep{2019A&A...622A..17S}; precisely the objects in which feedback is required the most. Treating $f_{\mathrm{RL}}$ as a proxy for the AGN jet duty cycle, the result implies a clear connection between the fuelling of the SMBH and {the state of} its surrounding environment. Yet, questions still remain regarding the nature of this relationship. Principally, if it is AGN feedback that links the quenching of star formation to the prevalence of the jets, then the energies involved must be quantified to understand how this process occurs. Additionally, the $f_{\mathrm{RL}}$ cannot separate between objects with rapid versus long-period duty-cycles; this distinction is important, considering that such objects will have vastly different duration active jet phases, and will thus differ in their feedback efficiencies \citep{2018MNRAS.480.5286Y}.

Valuable constraints towards AGN feedback mechanisms can be derived by examining the distribution of AGN jet kinetic powers and lifetimes. For a single AGN outburst, its total energy output can be found by integrating the instantaneous jet kinetic power (the jet power herein) over the the duration of the outburst (the jet lifetime herein). Considering that the jets are ultimately produced by accretion onto the SMBH \citep{1969Natur.223..690L}, we can expect each of these jet properties to be set by the conditions of and around the black hole; namely mass accretion rate and spin for jet power \citep{2009ApJ...696L..32D,2016MNRAS.458L..24D,2021MNRAS.500..215D}, and the underlying accretion mechanism for the jet outburst \citep{2011ApJ...737...26N,2013MNRAS.434..606G}. It follows that the distribution in jet powers and lifetimes not only constrains the energies released by AGNs, but also begins to probe the underlying mechanisms responsible for the production and fuelling of the jets. Considering also the mass-dependence of the AGN jet duty cycle, reconciling the global properties of the jets together with those of their environment are an important step towards linking AGN fuelling and feedback. 

Throughout the decades, a variety of techniques have been presented to measure the jet powers and lifetimes of radio-loud AGNs. The diffuse lobes inflated by the jets contain synchrotron-emitting plasma that forms an approximately power-law spectrum at radio frequencies. The energy-dependent synchrotron loss rate allows for a derivation of the plasma age based on the observed spectral steepening in the radio spectrum \citep[e.g.,][]{2007A&A...470..875P,2011A&A...526A.148M,2016A&A...585A..29B,2020PASA...37...37D}, and is referred to in the literature as the `spectral ageing' method \citep[e.g.,][]{1985ApJ...291...52M}. A shortcoming of this method is the requirement of knowledge of the lobe magnetic field strength, which in turn requires knowledge of the equipartition factor and the pressure in the ambient environment. By making several assumptions regarding the lobes, e.g. their composition and minimum energy, inferences of the jet power can be made based on their observed radio luminosity \citep{1999MNRAS.309.1017W}. Comparisons of these estimates to more sophisticated methods shows a general agreement \citep[e.g., see][]{2017MNRAS.467.1586I,2018MNRAS.474.3361T}, however with an appreciable level of scatter in part due to the departure from the underlying assumptions, but also due to the confounding impacts of the source age and environment \citep{2013ApJ...767...12G,2013MNRAS.430..174H}. Cavities in the hot X-ray emitting intracluster gas allow for robust estimates to be made of the $p dV$ work done by the expanding lobes; with an assumption about the source age (discussed above), inferences of the jet power can then be made. However, this technique is generally limited to low redshifts ($z \lesssim 0.5$) as the trade-off between resolution and surface brightness sensitivity in X-ray observations prevents measurements of the shape of the gas density profile beyond the cluster core \citep[e.g.,][]{2020MNRAS.493.5181T}.

Dynamical model-based methods have increasingly proved an attractive alternative for measuring the intrinsic properties of AGN jets. The current generation of analytical models (e.g.,\citealt{2018MNRAS.473.4179T,2018MNRAS.475.2768H,2023MNRAS.518..945T}; see \citealt{2023Galax..11...87T} for a review) match the evolutionary histories predicted by modern relativistic hydrodynamic simulations \citep[e.g.,][]{2013MNRAS.430..174H,2014MNRAS.443.1482H,2016MNRAS.461.2025E,2018MNRAS.480.5286Y,2019MNRAS.490.5807E,2021MNRAS.508.5239Y} whilst including a full treatment of radiative and expansion losses in their synchrotron emissivity calculations. The energetics of radio-loud AGNs can be constrained by comparing simulated attributes of synthetic radio sources (e.g., size and luminosity) against those for the observed objects. This comparison is typically done through a Bayesian parameter inversion \citep[e.g., see][]{2015ApJ...806...59T,2018MNRAS.474.3361T,2018MNRAS.476.2522T,2020MNRAS.491.5015M,2020MNRAS.499.3660T,2022MNRAS.514.3466Q}, allowing the intrinsic radio source parameters to be fitted. Such methods have the ability to simultaneously constrain the jet power and active lifetime of a radio source, in addition to the lobe magnetic field strength \citep{2018MNRAS.474.3361T} or host redshift \citep{2020MNRAS.499.3660T}, based on its angular size and multi-frequency radio spectrum. \citet{2022MNRAS.514.3466Q} showed that high-resolution radio images at two frequencies (e.g., 151\,MHz and 1.4\,GHz; typical of large-sky surveys) can provide sufficient constraining power in a Bayesian parameter inversion to fit at least four intrinsic jet parameters including the duration of the remnant phase, $t_{\rm rem}$. However, such parameter inversions rely on the ability of the underlying model to capture the necessary physics needed to explain the observation. For example, the analytical model of \citet{2023MNRAS.518..945T} does not currently capture processes such as jet precession or a bending of the jet, and may therefore be inapplicable for certain active radio sources; on the other hand, these issues are less critical for the remnant phase as regions of enhanced emission associated with the jets will rapidly fade due to adiabatic expansion and radiative loss mechanisms (e.g., see \citealt{2023Galax..11...74M} for a review). 

Recently, \citet{2020A&A...638A..34J} selected an observationally-complete sample of radio galaxies from the {LOFAR Two-metre Sky Survey} \citep[LoTSS;][]{2019A&A...622A...1S,2022A&A...659A...1S}. Through the application of forward modelling facilitated by the \textit{Radio AGN in Semi-analytic Environments} \citep[RAiSE; see][]{2023MNRAS.518..945T} model, \citet{2020MNRAS.496.1706S} used this sample to constrain the intrinsic (or true) jet power and lifetime functions (see also \citealt{2019A&A...622A..12H} for a similar approach). Their approach \citep[i.e., that of][]{2020MNRAS.496.1706S} relied on the comparison between mock and observed distributions in flux density and angular size, as well as the constraints from the observed fraction of remnant and restarted radio sources. Their work demonstrated that power-law lifetime models were needed to reproduce the observed properties of their reference sample, unlike constant-age models which under-predict the fraction of remnant radio galaxies. 

In this work, we combine and extend the approaches taken by \citet{2020MNRAS.496.1706S} and \citet{2022MNRAS.514.3466Q} to quantify the intrinsic jet power and lifetime functions for a sample of remnant radio galaxies. Such objects are unique in that they allow for a direct measurement of the full duration of the jet outburst, thus offering tight constraints on the AGN jet lifetime function. First, we identify a sample of 79 {candidate} remnant radio galaxies brighter than $0.5$\,Jy, with angular sizes $\theta \geqslant 4'$, and within a redshift range of $0.02 <z <0.2$ (\cref{sec:remnant_sample}). Next, we adopt the method of \citet{2022MNRAS.514.3466Q} to fit their intrinsic parameters, allowing us to probe their observed jet power and active age distributions (\cref{sec:sample_energetics}). To deal with the inherent selection biases, we construct mock remnant catalogues based on an assumed jet power and lifetime function, and filter the resulting radio source population by our sample selection criteria (\cref{sec:mock_remnant_populations}). We test a grid of jet power and lifetime models, and compare their predictions to our observed sample statistics to constrain the true jet power and lifetime functions (\cref{sec:results}). We discuss our results in the context of the broader literature and conclude in \cref{sec:ch4_conclusion}. 

Throughout this work, we assume the $\Lambda$CDM concordance cosmology with $\Omega_M$ = 0.3, $\Omega_\Lambda=0.7$ and $H_0$ = 70\,km\,s$^{-1}$\,Mpc$^{-1}$ \citep{2016A&A...594A..13P}, and the J2000 epoch for the equatorial coordinate system. We express the natural logarithm as $\ln(x)$ and the base-10 logarithm as $\log(x)$.

\section{Low-redshift sample of remnants}
\label{sec:remnant_sample}

We compile a sample of remnant radio galaxies from a range of large-sky radio surveys to form the observational basis for the later analysis in this work (\cref{sec:parent_sample}). The radio galaxies in this parent sample are cross-matched with infrared identified host galaxies, which are then matched to obtain optical passband photometries and redshifts, and to infer host galaxy properties (\cref{sec:cross_matching}). We then use literature-established methods to select {candidate} remnant lobes from our parent radio galaxy sample (\cref{sec:remnant_classification}). Finally, following the method of \citet{2022MNRAS.514.3466Q}, we measure the observable attributes of each remnant lobe needed to constrain its energetics (\cref{sec:radio_source_attributes}).

\subsection{Radio galaxy selection}
\label{sec:parent_sample}

We outline the steps followed to select a parent sample of radio galaxies from existing large-sky surveys. We impose the selection criteria that the radio sources are brighter than $0.5$\,Jy at 154\,MHz (i.e., $S_{154}\geqslant0.5$\,Jy) and larger than four arcminutes in size ($\theta>4'$); this ensures the measured radio source attributes are not heavily degraded either by (relatively) poor surface brightness sensitivity or a lack of spatial sampling across the lobe.

\subsubsection{Radio survey availability}
\label{sec:survey_availability}

We consider sources selected by both the {GaLactic and Extragalactic All-sky MWA} \citep[GLEAM;][]{2015PASA...32...25W,2017MNRAS.464.1146H} survey and the {Rapid ASKAP Continuum Survey} \citep[RACS;][]{2020PASA...37...48M} to ensure radio images are present at two or more frequencies, whilst maximising the potential sample of remnant radio galaxies. These two surveys both cover the following declination range: $-85^\circ \leqslant \mathrm{Decl.} \leqslant 30^\circ$. We supplement these observations with images from other large-sky surveys with partial coverage of this region: the {TIFR GMRT Sky Survey} \citep[TGSS;][]{2017A&A...598A..78I}, the {NRAO VLA Sky Survey} \citep[NVSS;][]{1998AJ....115.1693C} and the {Very Large Array Sky Survey} \citep[VLASS;][]{2019arXiv190701981L}. The properties of these surveys are summarised in \cref{tab:all_sky_radio_surveys}. Importantly, the observational parameter space provided by the GLEAM survey, principally the low-frequency coverage and low spatial-resolution, minimises the deselection of aged remnant lobes. 

\begin{table}
\centering
\begin{tabular}{lcccc}
\hline
{Survey}   & {Frequency} & {Sensitivity} & {Resolution} & {Decl. limit}\\
& (MHz) & (mJy/beam) & ($''$) &  \\ 
\hline
GLEAM      & 72--231 & $\sim7$    & $120$  &  $\leqslant+30^\circ$\\ 
TGSS & 151 & $\sim5$    & $25$   &   $\geqslant-53^\circ$ \\ 
RACS-low & 887 & $\sim0.25$ & $\sim25$  & $\leqslant+30^\circ$ \\
NVSS  & 1400 & $\sim0.45$     & $45$  &  $\geqslant-40^\circ$ \\
VLASS    & 3000 & $\sim0.12$ & $3$ & $\geqslant-40^\circ$ \\
\hline
\end{tabular}%
\caption[All-sky radio survey properties]{Properties of the large-sky radio surveys used in this work. Each row represents a unique radio survey (column 1), for which the observing frequencies (column 2), root mean square sensitivities (column 3), angular resolutions (column 4), and declination limits (column 5) are shown.  }
\label{tab:all_sky_radio_surveys}
\end{table}

\subsubsection{Integrated flux density cut: $S_{154}\geqslant0.5$\,Jy}
\label{sec:flux_cut}

We begin the parent sample compilation by considering all radio components brighter than $S_{154}\geqslant0.5$\,Jy. We conduct this step using the GLEAM survey due primarily to its enhanced surface brightness sensitivity to large-scale structure. As a preliminary step, we incorporate the GLEAM~4~Jansky catalogue \citep[G4Jy;][]{2020PASA...37...17W,2020PASA...37...18W}, which offers a complete GLEAM-selected sample of radio galaxies brighter than 4\,Jy at 151\,MHz{\footnote{The central frequency of the GLEAM 139-170\,MHz band is at 154\,MHz compared to a central frequency of 151\,MHz across the entire survey.}}, and also offers robust host galaxy associations. Importantly, all sources in the G4Jy catalogue meet the integrated flux density criterion of our sample, and are thus filtered through to our candidate sample. To avoid any duplicates of these sources, we remove components comprising the radio galaxies present in the G4Jy sample from the GLEAM catalogue prior to further source selection. This is done using a cone search centered at the coordinates of the G4Jy centroid, setting the radius of the search equivalent to reported angular size. 

\begin{figure*}
    \centering
    \begin{subfigure}{0.498\textwidth}
        \centering \includegraphics[width=0.925\textwidth,trim={17.5 8.5 0 12.5},clip]{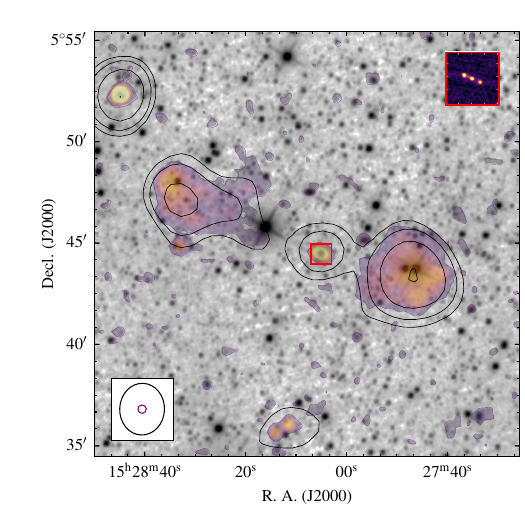}
        \caption{J1528+0544}
        \label{fig:J1528+0544}
    \end{subfigure}%
    ~ 
    \begin{subfigure}{0.498\textwidth}
        \centering
        \includegraphics[width=0.925\textwidth,trim={17.5 8.5 0 12.5},clip]{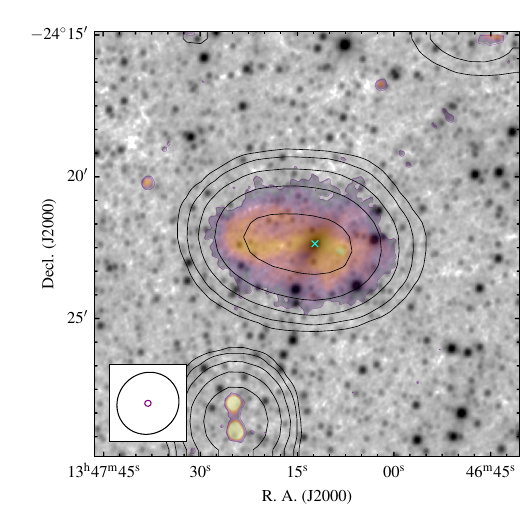}
        \caption{J1317-2422}
        \label{fig:J1317-2422}
    \end{subfigure}
    \caption{Example overlays used to manually inspect the GLEAM-selected radio sources. Grey-scale background images represent the AllWISE W1 bands. Each radio source is shown at 200\,MHz as seen by GLEAM (solid black contours), and as seen by RACS-low at 887\,MHz (semi-transparent images); the shape of their restoring beams are shown in the lower-left corner by the black and purple ellipses, respectively. For \cref{fig:J1528+0544}, a close-up view around the host galaxy coordinates, as seen at 3\,GHz by VLASS, is shown in the upper-right corner. For \cref{fig:J1317-2422}, the coordinates of the host galaxy are marked by the cyan cross. }
    \label{fig:example_remnants}
\end{figure*}

We now need to identify all GLEAM-selected radio sources with an integrated flux density of $0.5\leqslant S_{154} (\mathrm{Jy})\leqslant 4$. For each component in the GLEAM catalogue, we use the fitted $72$--$231$\,MHz spectral index ($\alpha_{72}^{231}$) and reference 200\,MHz integrated flux density ($S_{200\,\mathrm{fit}}$) to compute the 154\,MHz integrated flux density; we use this value over that measured from the $154$\,MHz GLEAM wide-band, in order to reduce the measurement uncertainty due to white noise. Components with flux densities of $S_{154}\geqslant0.5$\,Jy are added to our candidate sample. However, recognising that extended ($\theta>4'$) radio galaxies can appear as multiple components even in the GLEAM catalogue (e.g., see \cref{fig:example_remnants}), we must also consider radio sources for which the sum of their component flux densities exceeds our required threshold. To do this, we perform a cone search within the remaining GLEAM catalogue to internally cross match all components within a $\theta=18'$ radius; i.e., based on the largest angular size reported in the G4Jy catalogue. Component groups with a total flux density $S_{154}\geqslant0.5$\,Jy are also added to our candidate sample. 

Following this approach, we identify a total of 3071 candidate radio sources with $S_{154}\geqslant~0.5$\,Jy.

\subsubsection{Extended radio galaxy cut: $\theta \geqslant 4'$}
\label{sec:extended_cut}

We next need to identify which of our radio source candidates are: (1) extended radio galaxies larger than $\theta\geqslant4'$, and (2) are not comprised of components associated with unrelated radio sources. Considering the size of this sample is just on the limit of what can be handled manually, we perform this step largely through a visual inspection to ensure our classifications are robust. However, we perform two initial cuts to remove compact radio galaxies from our sample, as we now briefly discuss.

We use the {Tool for OPerations on Catalogues And Tables}~(\textsc{TOPCAT}) software, with a search radius of $\theta=2'$ (i.e., the size of the GLEAM resolution at 200\,MHz), to perform an internal cross match within the TGSS catalogue (for $\mathrm{Decl.} \geqslant -40^\circ$) to search for all isolated components within this radius. We then filter the TGSS components for sources with a integrated-to-peak flux density ratio less than $1.3$; i.e., definitely compact. These TGSS objects are cross-matched with the GLEAM catalogue, and we remove the associated GLEAM components from the sample. We perform the same internal cross match within the NVSS catalogue to search for isolated components in this survey, and cross-match these outputs with the GLEAM catalogue. By calculating a two-point spectral index between $154\,$MHz and $1.4$\,GHz, GLEAM components are excluded if the measured spectral index is greater than zero (i.e., brighter at higher frequencies). This is a conservative cut, considering that radio galaxy lobes should have a spectral index at least steeper than $\alpha=-0.5$, and is designed to exclude compact objects such as gigahertz peaked spectrum objects and blazars.

The remaining radio galaxy candidates are visually inspected to check that the components are associated with an actual radio galaxy and meet the angular size cut ($\theta\geqslant4'$). We use the coordinates of their radio centroid to create $0.5^\circ\times0.5^\circ$ postage stamps from the GLEAM, TGSS, RACS-low and VLASS radio images. For multi-component source candidates, we take the radio centroid as the mean of the individual component coordinates. Importantly, the (relative) high resolution and surface brightness sensitivity of the RACS-low radio imaging make it possible to inspect the radio morphologies of each candidate radio source in sufficient detail. We use the SAOImageDS9 \citep{2003ASPC..295..489J} software to manually measure the largest angular size (LAS) across the lobes. Sources are discarded if the measured LAS fails to reach the required angular size criterion. Objects are also discarded if their GLEAM component(s) are associated with unrelated sources. This is true for both single-component sources, i.e., where the singular GLEAM component is in fact multiple radio sources, as well as multi-component sources, i.e., where the GLEAM components themselves are revealed to be unrelated. These decisions are aided with infrared images indicating potential host galaxies locations (discussed in \cref{sec:cross_matching}), and also aided with the discarding of diffuse radio emission associated with nearby star-forming galaxies (SFGs).

Finally, following this approach, we arrive at a parent sample of 795 radio galaxies with $S_{154} > 0.5$\,Jy and $\theta\geqslant4'$. To prepare our sample for further analysis, we compile the integrated flux densities for each source from the {four} GLEAM wide-bands, RACS-low and NVSS (for $\mathrm{Decl.} \geqslant -40^\circ$).

\subsection{Local sub-halo environment}
\label{sec:cross_matching}

The properties of the environment within which the source expands need to be quantified to successfully model the evolutionary history of a radio galaxy. Specifically, we require estimates for the host galaxy redshift and stellar mass, as we briefly discuss. The RAiSE code uses semi-analytic galaxy evolution model outputs to identify plausible dark matter halo properties based on the measured stellar mass and cosmological evolution at the observed redshift. The median relationship between these variables is used to provide point estimates for the halo mass in this work \citep[see][]{2017Turnerthesis}. Meanwhile, the shape of the ambient gas density profile in RAiSE is informed by X-ray observations scaled to the properties of the chosen dark matter halo \citep{2015ApJ...806...59T,2023MNRAS.518..945T}. 

In this section, we cross match our radio galaxy sample to the {Wide-field Infrared Survey Explorer} (WISE) all-sky survey (\cref{sec:wise}), and use the WISE object IDs to automatically cross match our sample against various online data bases in order to obtain redshift measurements (\cref{sec:xmatch_redshift}) and optical passband photometries to calculate value-added products (\cref{sec:optical_gi}). Unless otherwise stated, all database cross matching outlined in this section is performed using the VizieR catalogue service\footnote{\url{https://vizier.cds.unistra.fr/}}.

\subsubsection{The AllWISE Data Release}
\label{sec:wise}

We perform our initial host galaxy cross match with respect to the AllWISE Data Release \citep{2014yCat.2328....0C} publicly available on VizieR\footnote{https://vizier.cds.unistra.fr/viz-bin/VizieR?-source=II/328}. We choose this survey considering it covers the full declination range occupied by our sample, and also probes a wavelength range well-suited to the detection of dust-obscured host galaxies. Additionally, its most sensitive passband (W1, $\lambda = 3.4\,\mu$m) has a high angular resolution of $\sim4''$, comparable to that of VLASS. We create similar postage stamps to those discussed in \cref{sec:extended_cut} for the $\lambda=3.4\,\mu$m band of the WISE survey \citep{2010AJ....140.1868W}. These images were used to identify the coordinates of the likely host galaxy based on the orientation of the radio emission with respect to the galactic environment. For radio sources with a clear radio core and/or central jet, we took the host galaxy as that which coincided with the origin of the radio emission. We used the highest resolution radio data available, which at minimum was provided by RACS-low, however in many cases was improved on by VLASS imaging. For radio sources without a clear radio core, we recorded the likely coordinate of the host galaxy based on the orientation of the diffuse lobe plasma with respect to the surrounding galaxies. We obtain the WISE object ID for the candidate host galaxies identified using the WISE W1 images (see \cref{fig:example_remnants}). 

{Host galaxy identification is often challenging for remnant radio galaxies \citep[see e.g.,][]{2017A&A...606A..98B,2021PASA...38....8Q}, however, we did not encounter any cases where the host galaxy association was ambiguous. This is expected due to our high flux density cut at radio frequencies relative to the sensitivity of the WISE survey; i.e., the radio--mid-infrared relationship predicts the WISE W3 flux density of the most AGN-dominated galaxies \citep[cf.][]{Mingo+2016} is orders of magnitude brighter than the 5$\sigma$ point source sensitivity for WISE \citep[$S_\text{W3} \gg 1$\,mJy;][]{2010AJ....140.1868W}. Meanwhile, the large angular size cut reduces the likelihood of a high-redshift host galaxy given the rarity of giant radio galaxies (e.g., a size $>$$1.5$\,Mpc at redshift $z \geqslant 0.5$). These two factors, together, ensure the host galaxy is present in the WISE image and has a low chance of being spatially coincident with a more proximate galaxy.}


\subsubsection{Host galaxy redshifts}
\label{sec:xmatch_redshift}

\begin{figure*}
    \centering
    \includegraphics[trim={0 15 0 0},clip]{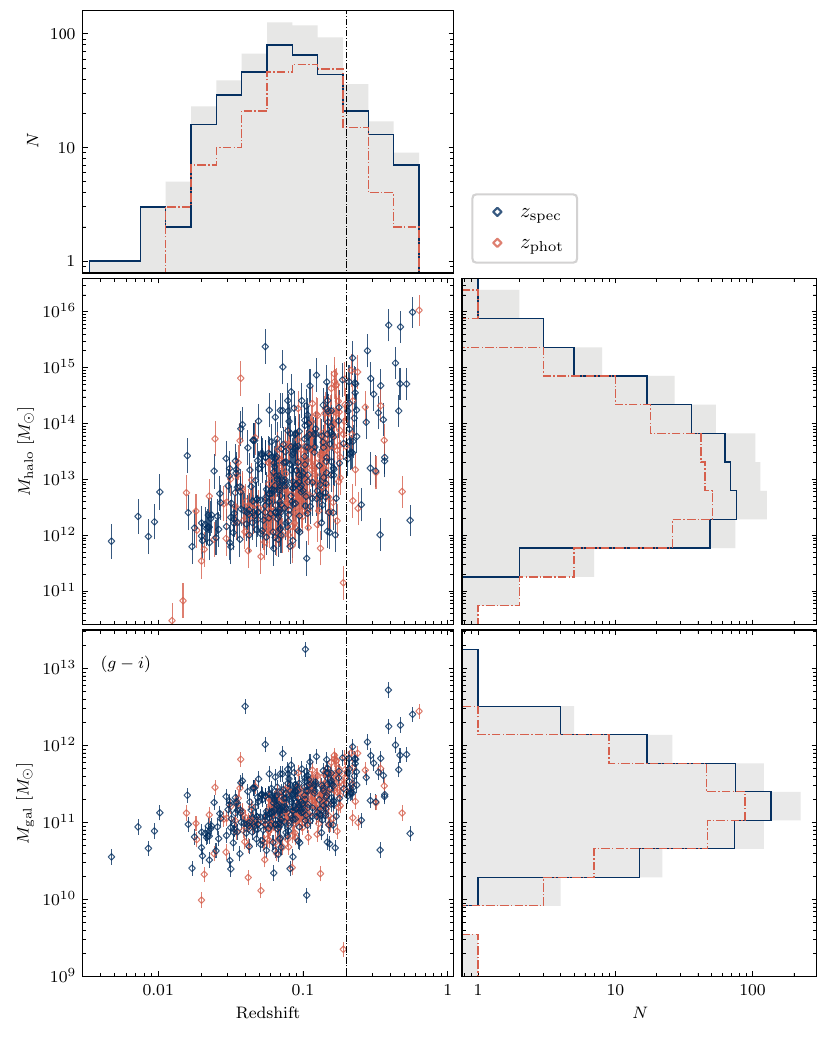}
    \caption{The redshift, stellar and halo mass distributions for our parent radio galaxy sample. The two scatter plots show the derived halo masses (upper panel) and stellar masses (lower panel) as a function of redshift. Spectroscopic and photometric redshift are denoted using blue and orange colors, respectively. Each histogram shows the one-dimensional distribution in the parameter adjacent to the plot. The grey bars denote the full sample (spectroscopic and photometric). }
    \label{fig:z_M_H_dist}
\end{figure*}

Redshift information is obtained for our radio sources using the astroquery \citep{2019AJ....157...98G} \textsc{python} module to cross match our host galaxy coordinates with the NASA/IPAC Extragalactic Database\footnote{\url{https://ned.ipac.caltech.edu/}} (NED). For each radio source, we perform a $\theta=2'$ cone search around the coordinates of the host galaxy, and filter the outputs for objects with catalogued redshifts in NED. Through this approach, redshifts are found for 540 out of the original 795 host galaxies. We also record the classification of each redshift type, whether the redshifts are spectroscopic or photometric, based on the original references for each catalogued redshift. Overall, our sample was broken down into 329 spectroscopic, and 211 photometric redshifts. Our redshift distribution is shown in \cref{fig:z_M_H_dist}. As a sample cut, we limit the redshift distribution to $0.02 < z < 0.2$, to reduce the contamination from sources with photometric redshifts (but see also discussion in \cref{sec:input_parameter_space}). At $z\geqslant0.2$, our photometric redshifts become more numerous than spectroscopic which, considering the redshift is needed to perform the parameter inversion, may introduce an unwanted systematic into the fitted energetics. 

Since these redshifts are heterogeneously-selected, we can not comment on the redshift completeness of our sample. However, we clarify that the method undertaken for this work is not reliant upon volume-complete radio galaxy samples. In \cref{sec:mock_remnant_populations}, the predictions made by forward modelling a given jet power and lifetime model are made specific to the intrinsic property distributions of the selected remnant sample. This includes the redshift distribution, which in principle means that the results should not be biased by redshift incompleteness. {We caution that a small bias may be introduced if certain host galaxies (e.g., lower stellar masses) are preferentially excluded from our analysis due to this redshift incompleteness; however, we explicitly control for the properties of our sample in \cref{sec:controlling}, so our findings will be robust for the bright end of the stellar mass and radio luminosity functions probed in this work.}

\subsubsection{Host galaxy stellar masses}
\label{sec:optical_gi}

We derive the stellar mass of each host galaxy based on the empirical relationship between the rest-frame $(g-i)$ colour and mass-to-light ratio ($M_*/L_i$) \citep{2011MNRAS.418.1587T}. The stellar mass is given by their linear best fit (e.g., see their equation~8):
\begin{equation}
\label{eqn:stellar_mass}
    \log [M_*/M_\odot] = -0.68 + 0.70(g-i) - 0.4 m_i - 2\log \Big(\frac{d_\mathrm{L}}{10\,\mathrm{pc}}\Big),
\end{equation}
where $m_i$ is the apparent magnitude in the photometric $i$-band (AB system), $d_\mathrm{L}$ is the luminosity distance, and the $1\sigma$ uncertainty corresponding to the derived $M_*/L_i$ is $\sim0.1$\,dex. 

However, no single optical survey provides coverage of our entire radio galaxy sample. As such, we resort to compiling measurements from various surveys, specifically the {Panoramic Survey Telescope and Rapid Response System} \citep[Pan-STARRS1;][]{2016arXiv161205560C}, the {Sloan Digital Sky Survey} Data Release 12 \citep[SDSS DR12; ][]{2015ApJS..219...12A}, and the {Skymapper Southern Sky Survey} \citep[SMSS;][]{2018PASA...35...10W}. Using the coordinates of the WISE object IDs obtained above, we perform a cone search of each database using a $\theta=5''$ search radius to retrieve the best positional match for each object. For each survey, we obtain the AB magnitudes in the $g$ and $i$ passbands. For regions below a declination of $-40^\circ$, photometries are taken from SMSS. Photometries are otherwise taken from PanSTARRS, however for a small fraction of these objects ($\sim1\%$), the PanSTARRS photometry is either missing in one of the passbands, or has a poorer signal to noise ratio than that in SDSS. In these two cases, optical magnitudes are sourced from SDSS instead. 

In order to minimise any potential systematic biases that could lead to inconsistent stellar-mass estimates from multiple surveys for the same galaxy, we apply corrections to the SMSS and SDSS $g$ and $i$ magnitudes to align them with PanSTARRS, which serves as the reference due to its overlap with both surveys. The galaxies in our sample that appear in the overlapping region with PanSTARRS are used for independent linear corrections of the $g$ and $i$ bands in each survey.

\begin{figure*}
    \centering
    \includegraphics{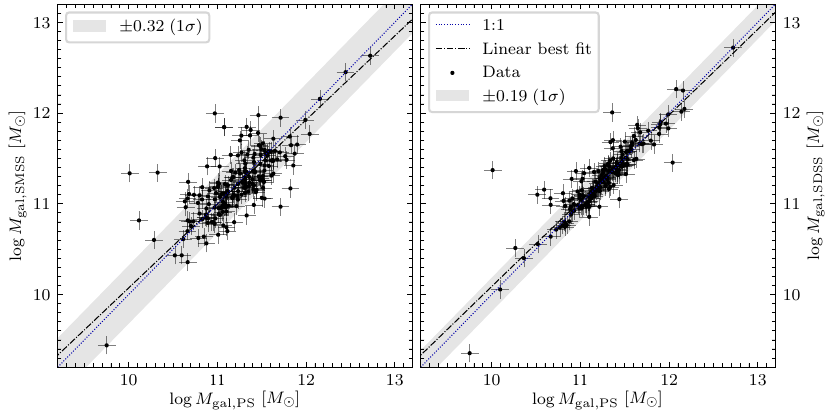}
    \caption{Consistency check for stellar masses derived using the $(g-i)$ colors from the PanSTARRS, SMSS and SDSS surveys. In each panel the PanSTARRS-based mass estimate (horizontal axis) is compared to the SMSS-based mass estimate (left panel) and the SDSS-based mass estimate (right panel). The standard deviation of the differences calculated between the two mass estimates is shown in the label for each panel, and represented by the grey-shaded region.}
    \label{fig:stellar_mass_comparison}
\end{figure*}

In \cref{fig:stellar_mass_comparison}, we compare the stellar mass estimates (\cref{eqn:stellar_mass}) of each survey for sources in the overlapping region with PanSTARRS. For the comparison of both the SMSS and SDSS surveys against PanSTARRS, we find the linear best fit to the data and compute the standard deviation of the residual. Reassuringly, our results demonstrate good agreement between all three stellar mass estimates. While the scatter is clearly larger for the SMSS-PanSTARRS comparison, both comparisons show a near 1:1 relationship, and thus do not appear to be systematically offset. We assume the same holds true for the entire sample. To account for the noisier SMSS data, we assign a $0.32$\,dex uncertainty (e.g., the $1\sigma$ deviation of the scatter) on the SMSS-derived stellar masses.

Our stellar mass distribution, and corresponding halo mass distribution derived from semi-analytic galaxy evolution model outputs \citep{2015ApJ...806...59T}, is shown in \cref{fig:z_M_H_dist}. The vast majority of galaxies have stellar masses between $10^{10.5}$ and $10^{12}$\,M$_\odot$, however, a few have extreme values fitted. The stellar mass of these galaxies is likely poorly constrained with two wavelength bands, however, as the mass is solely used as a prior to inform the shape of the host galaxy ambient medium, we simply assume semi-plausible values (either $10^{10.5}$ or $10^{12}$\,M$_\odot$) for these objects.

\subsection{Candidate remnant radio galaxies}
\label{sec:remnant_classification}

We {identify candidate} remnant radio galaxies based on the findings of \citet{2017A&A...606A..98B}, \citet{2018MNRAS.475.4557M}, \citet{2020A&A...638A..34J} and \citet{2021PASA...38....8Q}, who show that remnants can display a wide range of observed properties. In \cref{sec:sample_energetics}, we will apply forward modelling to directly constrain the energetics of the sample{, including the off-time, $t_\text{rem}$ \citep[see][]{2022MNRAS.514.3466Q}, to confirm the remnant status of these radio galaxies}. {For this reason, the primary aim of this initial remnant classification is to exclude most (and preferably all) active radio lobes while maintaining an otherwise unbiased sample. As such, we consider several remnant classification pathways to minimise the preferential deselection of any remnant sub-populations.}
{Importantly, the remnant lobes associated with restarted sources can be modelled through a RAiSE-based parameter inversion provided the two epochs of radio activity do not overlap spatially. With this in mind, we adapt criteria for remnant and restarted radio source selection -- applied to the cutouts created in \cref{sec:extended_cut} -- to construct a sample of candidate radio galaxies with associated remnant lobes, as we now describe.}

{We define four metrics upon which we base our candidate remnant radio lobe selection:
\begin{itemize}
    \item `Absent radio core' metric \citep[][]{2021PASA...38....8Q}. The (coordinates of the) host galaxy has no radio emission detected at the $3\sigma$ level, where $\sigma$ is the local root mean square (rms) noise. We analyse the RACS-low data across the entire sample, but additionally include VLASS data for $\text{Decl.}\geqslant-40^{\circ}$ sources.
    \item `Absent compact features' metric, as for the radio core criterion above, but using the VLASS and/or RACS-low data to identify the presence of jets and/or hotspots in candidate lobes.
    \item `Low surface brightness' metric based on the ratio of the GLEAM 154\,MHz integrated flux density and the projected area of the lobe from the higher-resolution RACS-low data (i.e., average surface brightness); we use the \textsc{polygon\_flux}\footnote{\url{https://github.com/nhurleywalker/polygon-flux}} code \citep{2019PASA...36...48H} to trace the $3\sigma$ footprint of each radio source. The RAiSE model confirms/informs a surface brightness threshold of $50\,$mJy/arcmin$^{-2}$ \citep[cf.][]{2021A&A...653A.110J} below which the lobes of old, high jet power radio sources -- expected for our angular size cut -- are confidently inactive based on their rapid fading upon the cessation of jet activity \citep[cf. Fig. 4 of][]{2018MNRAS.476.2522T}\footnote{The large projected area of old, high jet power sources ensures their average surface brightness rapidly falls below that of lower-powered active sources.}.
    \item `Ultra-steep spectrum' metric based on the spectral index of the candidate lobe (excluding emission from the core region if present). The spectrum is fitted using a power-law (between the lowest and highest frequency available) with a spectral index steeper than $\alpha<-1.2$ suggesting remnant lobes; i.e., steeper than expected for an active lobe with a typical injection spectral index of $\alpha_\text{inj} \geqslant -0.7$ \citep[e.g.,][]{2018MNRAS.474.3361T}. 
\end{itemize}}

{We are now able to select candidate remnant radio galaxies from our parent sample using combinations of these metrics.
Firstly, we classify all radio galaxies satisfying the `absent radio core' metric as candidate remnants. However, \citet{2021A&A...653A.110J} proposed that faint radio emission from the core, perhaps due to the jets not completely shutting off or star formation at lower redshifts, can still coincide with remnant radio lobes. For radio sources with a core detection, we classify sources as remnant candidates if their lobes satisfy both the `absent compact features' and `low surface brightness' metrics \citep[cf. the `morphological' criterion of][]{2017A&A...606A..98B}. This criterion is supported by our selection of the known remnant radio galaxy associated with NGC~1534 \citep{2015MNRAS.447.2468H,2019PASA...36...16D}, which has unambiguously remnant lobes, yet also shows weak radio emission at the host galaxy. Finally, in line with the restarted radio galaxy classification techniques of \citet{2020A&A...638A..34J}, we identify candidate remnant lobes when a clear radio core is observed if the lobes satisfy all three of the other metrics: the `absent compact features', `low surface brightness' and `ultra-steep spectrum' metrics. }

{Following these methods, we obtain a final sample of 79 candidate remnant radio galaxies within $0.02< z < 0.2$. The RAiSE parameter inversion in \cref{sec:sample_energetics} confidently fits non-zero off-times for 68 of these objects (2$\sigma$ level; see \cref{sec:appendix}) providing an independent validation of our candidate selection technique. The eleven objects not confidently modelled as remnant lobes include five objects with no radio core detection in RACS-low (though two have a weak detection in VLASS). We note the findings from our subsequent analysis are unaffected by the inclusion (or not) of the remaining six less confident remnant candidates as, in particular, their fitted jet powers and active ages lie at the peak of the distribution for both these parameters (see \cref{sec:sample_energetics}).}

\subsection{Measuring the radio source attributes}
\label{sec:radio_source_attributes}

We need to measure a set of attributes from the observed radio images of our selected remnant radio galaxies in order to perform a Bayesian parameter inversion for their energetics (\cref{sec:sample_energetics}). In the RAiSE framework, these observed radio source attributes are compared to those extracted from synthetic radio sources using a maximum likelihood algorithm. It follows, then, that the quality and number of measured radio source attributes (unique constraints) must be adequate to constrain the underlying parameters. For this work, we implement the result of \citet{2022MNRAS.514.3466Q}, who showed that attributes related to the observed surface brightness distribution of remnant lobes help to constrain the duration of the remnant phase. 


With this in mind, and given the quality of the radio data available to us (see \cref{sec:survey_availability}), the following radio source attributes are of particular interest to us:
\setitemize{topsep=3pt,parsep=0pt,itemsep=0pt,leftmargin=3\parindent,itemindent=-1.5\parindent}
\begin{itemize}
    \item a reference radio luminosity;
    \item a largest linear size;
    \item the two extent attributes, which describe the lobe surface brightness distribution \citep[e.g.,][]{2022MNRAS.514.3466Q};
    \item a low frequency spectral index; and
    \item a measure of spectral curvature.
\end{itemize}
We define the reference radio luminosity, $L_\nu$, as the 154\,MHz integrated flux density measured for each remnant. The decision to measure this attribute using GLEAM data is motivated by the sensitivity to low surface brightness emission and relatively low calibration uncertainty \citep[e.g., approximately 8\%;][]{2017MNRAS.464.1146H}. However, for three of our selected remnants, blending with unrelated components resulted in an uncertain measurement of their $S_{154}$. For these sources, we measured their reference radio luminosity at 887\,MHz using RACS-low data (which can be equally well predicted by RAiSE). 

To measure the largest linear size and the two extent attributes, we use the method of \citet{2022MNRAS.514.3466Q}, who parameterise these attributes by fitting the lobe surface brightness profiles with skewed Gaussian functions (e.g., see their section~4.2.2). As part of their analysis, they demonstrated that the extent attributes are mostly affected by the spectral ageing of the oldest plasma, meaning that higher-frequency measurements are more constraining. We therefore measure the extent attributes at 887\,MHz, considering that the RACS-low observations provide the optimal trade-off between frequency, sensitivity, and spatial resolution; for example, while NVSS probes higher frequencies, the restoring beam is also larger, and thus does not offer much constraining power here.

The low-frequency spectral index, $\alpha_{\mathrm{low}}$, is measured using the {four} GLEAM 30--60\,MHz-wide bands (centres ranging between 88\,MHz and 200\,MHz). In the ideal case, the spectral index is measured across all {four} bands. However, for a handful of cases, blending from unrelated components (which becomes increasingly worse towards lower frequencies) makes this measurement problematic. For two such sources, the low-frequency spectral index was measured between 154\,MHz and 200\,MHz. To derive the spectral index, spectral data points were fit in log-log space (flux density and frequency) using a simple linear fit. 

Finally, to obtain the spectral curvature, SPC, we measured a higher-frequency two-point spectral index, $\alpha_{\mathrm{high}}$, and calculated the difference from $\alpha_{\mathrm{low}}$ \citep[see][]{2022MNRAS.514.3466Q}. To measure this attribute, we used one of two ways. If the radio source was detected by NVSS, $\alpha_{\mathrm{high}}$ was calculated between 887\,MHz and 1.4\,GHz. Consequently, if the radio source fell below the NVSS declination limit ($\text{Decl. = }-40^\circ$), or above the declination limit but was undetected by NVSS, we calculated $\alpha_{\mathrm{high}}$ between $200$\,MHz and $887$\,MHz. We found that all sources detected by NVSS were also detected by RACS-low, which is unsurprising considering the deeper and comparatively lower-frequency observations provided by RACS-low.

With this approach, the radio source attributes are measured for each remnant in our sample. 
Using these attributes, we now seek to model the evolutionary histories of our remnants, and thus constrain their energetics. This is discussed in the following section.

\section{Constraining the sample energetics}
\label{sec:sample_energetics}

The energetics of an observed radio source can be constrained via a RAiSE-based parameter inversion -- a method that exploits the ability of the RAiSE model to convert from intrinsic parameter space into observed attribute space. In the following section, we describe the adoption of this method to quantify the energetics for the sample of remnants compiled in the previous section. Briefly summarised, synthetic radio sources are generated across a multi-dimensional grid of input parameters that are expected to influence the observed properties (\cref{sec:underlying_raise_sim}). The parameter space is inverted by comparing synthetic attributes from the simulated surface brightness maps to their observed counterparts, whilst carefully modelling the redshift and survey limitations of each radio source. (\cref{sec:measuring_the_source_energetics}). We use these results to explore the observed distributions in the fitted parameters (\cref{sec:observed_jet_age_dist}).

\subsection{Creating the RAiSE simulation grid}
\label{sec:underlying_raise_sim}
 
The RAiSE simulations that underlie this work play a key role in both estimating the energetics of each observed remnant (\cref{sec:maxlikelihood_gridsearch}), and generating mock remnant populations required later in this work (\cref{sec:mock_remnant_populations}). We assume these remnants have a \citealt{1974MNRAS.167P..31F} Type-II morphology in their active phase, given this class is associated with higher jet powers {\citep[e.g.,][]{2009AN....330..184B}}, and thus {the class that will primarily be included in our sample due to the combination of our high angular size ($\theta > 4'$) and flux density ($S_{154} > 0.5$\,Jy) cuts; this is confirmed in \cref{sec:observed_jet_age_dist} by the high jet powers constrained for the majority of our sample\footnote{\citet[][their Fig. 3]{2020MNRAS.496.1706S} found the FR-I and -II models of \citet{2015ApJ...806...59T} predict comparable jet powers and ages in a parameter inversion; we do not expect any significant errors in our analysis if a small number of remnants are in fact of FR-I morphology.}}. Towards both of these aims, we must ensure that the input parameter space covers the full breadth of plausible values, and that the grid resolution in each dimension is at least comparable to the associated fitting uncertainty. Below, a detailed setup of the RAiSE simulation is outlined.

\subsubsection{Initial setup}
\label{sec:basic_setup}
To predict the evolutionary histories of a radio source, we use the \citet{2023MNRAS.518..945T} `jet+lobe' RAiSE model, for which the associated \textsc{python} code is publicly available on GitHub\footnote{\url{github.com/rossjturner/RAiSEHD}}. Radio sources are simulated using the \texttt{RAiSE\_run} function, which generates spatially-resolved images of the lobe synchrotron radiation for an set of model inputs. Two such inputs, \texttt{frequency} and \texttt{resolution}, control the properties of the output images. For computational efficiency, we use \texttt{resolution=poor}, which sets the number of Lagrangian particles used in the surface brightness calculation equal to 1,792,000. To match the combined frequency coverage provided by the GLEAM, RACS-low, and NVSS radio surveys, the synchrotron radiation is synthesised at 88, 154, 200, 887 and 1400\,MHz. The inputs related to the dynamics and energetics of the lobes represent dynamically relevant parameters that we will need to fit for each source and are discussed in the following section. The initial axis ratio could arguably be varied, but due to computational constraints, we assume a fixed value of $A_0=2.5$ consistent with observations of 3CRR \citet{1974MNRAS.167P..31F} Type-II radio galaxies \citep{2018MNRAS.474.3361T}; crucially, \citet[][their section 6.5]{2018MNRAS.473.4179T} show that this parameter has a very weak influence on the fitted energetics. Other input parameters for the jets/lobes, e.g. those describing the internal structure of the jet, are set to default values that have been calibrated against hydrodynamical simulations \citep[see Table~1 of][]{2023MNRAS.518..945T}. Finally, in these simulations we underlying the radio source environments based on a halo mass, which places the radio source at the centre of a spherically-symmetric, double-beta gas density profile \citep{2015ApJ...806...59T}.

\subsubsection{The input parameter space}
\label{sec:input_parameter_space}

The RAiSE-based parameter inversion uses attributes measured for an observed radio source to infer its intrinsic  parameters. In this approach, a hyper-dimensional grid of order $k$ is required to fit $k$ parameters, where each dimension is a list of uniformly-gridded values (in either log- or linear-space). Our main goal is to fit the energetics of each radio source, meaning we need to fit the jet power and active age{; $Q$ and $t_{\mathrm{on}}$ respectively}. However, we also want to account for other intrinsic properties that will have an influence on the observed radio source properties, thus ensuring that the fitted energetics are not systematically incorrect. These properties are {\citep[see][]{2022MNRAS.514.3466Q}:%
\setitemize{topsep=3pt,parsep=0pt,itemsep=0pt,leftmargin=3\parindent,itemindent=-1.5\parindent}
\begin{itemize}
    \item remnant ratio, $R_{\mathrm{rem}} = t_{\mathrm{rem}}/(t_{\mathrm{on}} + t_{\mathrm{rem}})$, where $t_{\mathrm{rem}}$ is the time spent in a remnant phase upon observation;
    \item halo mass (including dark and baryonic matter), $M_\text{halo}$;
    \item equipartition factor (ratio of energy in the magnetic field and particles) of the lobes, $q$; and
    \item  energy injection index, $s$, where the injection spectral index is $\alpha_\text{inj} = (1 - s)/2$.
\end{itemize}
The} intrinsic parameter space is thus created by uniformly gridding each aforementioned parameter between an upper/lower limit; see \cref{tab:grid_properties} for details. This results in a $k=6$ dimensional grid of mock radio sources.

\begin{table*}
    \centering
    \caption{The properties of the parameter space upon which the hyper-dimensional RAiSE simulation grid is created. Each row represents one of the six dimensions of the simulation grid, which is gridded between an upper/lower limit (column 3) using a fixed resolution (column 5), and comprises a total of $n$ values (column 6). This parameter space results in a total of $8.63\times10^{7}$ synthetic radio sources. }
    \begin{tabular}{cccccc}
    \hline
         Input parameter & Symbol & Value(s) & Unit & Resolution & $n$ \\\hline
         Jet power & $Q$ & $10^{36.5}$ - $10^{40}$ & W & $0.1$\,dex & $36$ \\
         Active age & $t_{\mathrm{on}}$ & $10^{6.5}$ - $10^{8.5}$ & Myr & $0.05$\,dex & $41$ \\
         Remnant ratio & $R_{\mathrm{rem}}$ & $0$ - $0.7$ & - & $0.025$ & $29$ \\
         Halo mass & $M_\mathrm{halo}$ & $10^{11.8}$ - $10^{15}$ & $\text{M}_\odot$ & $0.2$\,dex & $16$ \\
         Equipartition factor & $q$ & $10^{-2.7}$ - $10^{-0.7}$ & - & $0.1$\,dex & $21$ \\
         Injection index & $s$ & $2.2$ - $2.7$ & - & 0.1 & $6$ \\\hline
    \end{tabular}
    \label{tab:grid_properties}
\end{table*}

Ideally, to fit the intrinsic parameters for our remnant sample, a unique RAiSE simulation grid would be created at the redshift of each remnant. However, the RAiSE model outputs are only a weak function of redshift: notably, radiative losses from the inverse-Compton upscattering of CMB photons depends on redshift as $(1~+~z)^4$. Considering that the redshifts of our observed remnant sample is low ($z < 0.2$) we can reasonably approximate this loss mechanism with a simulation at a single redshift, $z=0.1$. This reduces the dimensionality of our parameter space thereby reducing the total computation time by a factor of 79 (our sample size). The observable attributes are thus calculated for the observed redshift of each remnant by scaling the RAiSE simulation outputs (run at $z=0.1$) prior to convolution with the relevant survey beam.

Following this approach, the RAiSE simulation grid is constructed by simulating the radio source evolutionary histories for each point in the input ($k=6$ dimensional) parameter space. Each point contains spatially-resolved maps of the synchrotron radiation expected at the frequencies specified in \cref{sec:basic_setup}. The integrated radio luminosities (for each input frequency) and the true physical size (measured from hotspot to hotspot) are also are calculated by default for each RAiSE model.  

\subsection{Modelling the intrinsic parameters}
\label{sec:measuring_the_source_energetics}
The energetics of each remnant can now be estimated via a parameter inversion using the RAiSE model as the theoretical expectation. To do this, the synthetic images must be scaled to match the redshift of the remnant, then be degraded to match the sensitivity and angular resolution of the radio observations; the radio source attributes are then measured for each synthetic image (\cref{sec:parameter_inversion}). The observed and synthetic radio source attributes are compared for each synthetic image using a maximum likelihood algorithm and weighted using Bayesian priors for properties of the host environment and lobe magnetic field. This gives probability density functions for the $k=6$ intrinsic parameters of each remnant in our sample, e.g., jet power, active age and off-time (\cref{sec:maxlikelihood_gridsearch}).

\subsubsection{Synthetic radio source attributes}
\label{sec:parameter_inversion}

The RAiSE simulations are run at a single redshift for computational efficiency, as discussed in \cref{sec:input_parameter_space}. These two-dimensional (2D) synchrotron-emission maps are recast into the frame of the observed radio source based on the measured redshift; in this process, we convert the pixel values of the output image from a luminosity (W/Hz) into a flux density (Jy), and similarly convert the scale of the pixel from a physical size (kpc) into an apparent size ($''$). The scaled outputs are convolved with a  2D Gaussian kernel with a scale and orientation matching the angular resolution of the relevant survey (e.g., to mimic a RACS-low observation, images at 887\,MHz are convolved with a circular beam of diameter $25''$). These convolved outputs are converted into synthetic surface brightness images (pixel values in units of Jy/beam) by dividing through by the area of the restoring beam. We note that the synthetic images do not include a random noise component as our sample is selected to have a good signal to noise ratio. Simulated radio source attributes, matching those measured for each remnant (\cref{sec:radio_source_attributes}), are extracted from the synthetic, survey-degraded surface brightness images \citep[see also][]{2022MNRAS.514.3466Q}. Following this approach, we obtain a set of attributes for the mock radio source generated by each combination of potential intrinsic radio galaxy properties -- these sets of attributes can be compared to those of the observed remnants in a parameter inversion.

\subsubsection{Bayesian parameter estimation}
\label{sec:maxlikelihood_gridsearch}

The energetics of each remnant are fitted using a maximum likelihood algorithm based on a comparison of the observed radio source attributes (\cref{sec:radio_source_attributes}) to the ($k=6$) hyper-dimensional grid of attributes for each combination of intrinsic parameters modelled using RAiSE (\cref{sec:parameter_inversion}). Following \citet{2022MNRAS.514.3466Q}, the Akaike information criterion (AIC) corresponding to each RAiSE model is calculated by comparing the observed and synthetic attributes via the likelihood function (e.g., their equation~9). The optimal model (or set of intrinsic parameters) is selected as that which minimises the AIC; the probability of each model is calculated relative to this best-fit. These probabilities only encode the agreement between simulation and observation, and do not reflect the Bayesian priors that exist for several parameters. For this work, the evaluated probabilities are multiplied through by two priors: the logarithmic equipartition factor, $\log q$, which is known to be approximately Gaussian-distributed \citep[$\mu\approx-1.73$, $\sigma\approx0.53$;][]{2018MNRAS.474.3361T}; and the halo mass, for which we obtained a measure of the mean and uncertainty in \cref{sec:cross_matching} based on the $(g-i)$ stellar mass.

Following this approach, the energetics of each remnant in our sample are constrained within the hyper-dimensional parameter space. The expectation and uncertainty in each of the $k=6$ parameters are derived from the marginal probability density functions (PDFs); i.e., the probabilities obtained sum over all but the parameter of interest. We take the 50th percentile of the PDF as our estimate of the best fit, and the 16 and 84th percentiles as our estimates of the 1$\sigma$ confidence interval. We examine the fitted properties of our remnant sample in the following section.

\subsection{Observed jet-power and lifetime distributions}
\label{sec:observed_jet_age_dist}

\begin{figure*}
    \centering
    \includegraphics{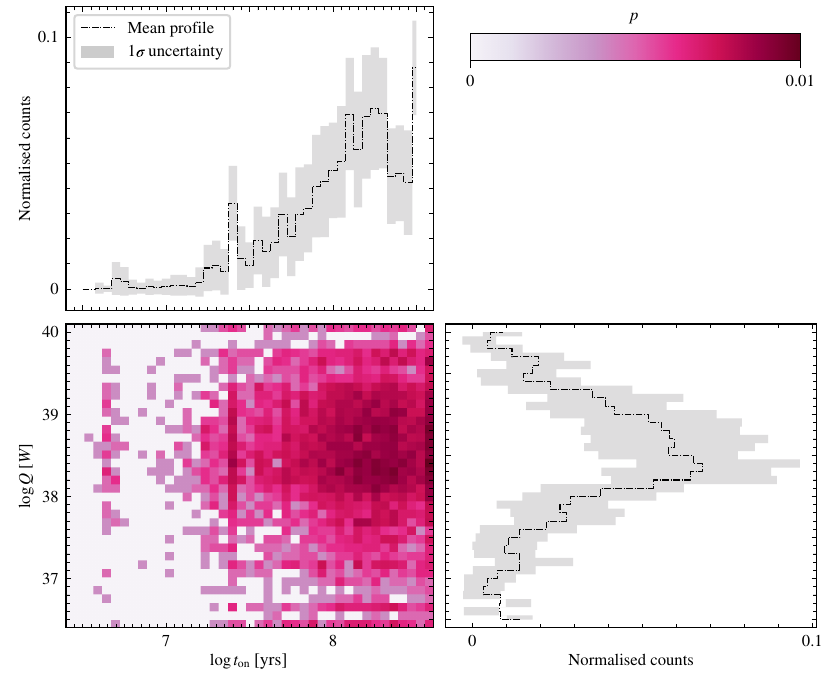}
    \caption{The observed marginal PDFs in jet power (lower right) and active age (upper left) fitted for our remnant sample. Monte Carlo simulations are used to build up the observed sample statistics in each parameter (grey shading). The two-dimensional grid (lower right) shows the observed joint PDF in $( Q, t_{\mathrm{on}})$-space. The colour-map indicates the mean probability density, $p$, of sources at each pixel.}
    \label{fig:fitted_sample_energetics}
\end{figure*}

We have now constrained the energetics of each observed remnant using the RAiSE-based parameter inversion, notably the jet kinetic power and duration of the active phase. We use Monte Carlo (MC) statistics to characterise the shape and uncertainty in the fitted jet power and lifetime distributions for our sample. Each MC realisation consists of $N=79$ jet power and active age estimates based on randomly sampling the probability density functions for each remnant. The resulting jet power and lifetime distributions for each realisation are then binned and normalised (area equal to unity) in a two-dimensional parameter space (i.e., in jet power and active age), where the bins are based on the gridded parameter space of the RAiSE simulation grid (\cref{tab:grid_properties}). We perform 1000 MC realisations to obtain an estimate for the expectation and uncertainty in the simulated counts for each bin in the jet power--active age, $( Q, t_{\mathrm{on}})$, parameter space. Our results are shown in \cref{fig:fitted_sample_energetics}.

Importantly, due to the sample selection bias, these distributions do not represent the true underlying statistics for these parameters. We can see that the observed jet power distribution peaks at $Q \approx 10^{38.5}$\,W, and shows an observed dearth of low-powered sources; this turn-over is likely due to Malmquist bias (see also \cref{sec:mock_remnant_populations} below) -- the preferential selection of bright objects in sensitivity-limited surveys. A similar effect, primarily due to the minimum angular size cut, can be seen for the active age distribution, which peaks at $t_{\mathrm{on}}\approx 10^{8.25}$\,yrs. 

In \cref{sec:mock_remnant_populations}, we will attempt to decode this sample selection bias by comparing mock remnant population with known jet power and lifetime functions to our observed distribution. To this end, it is worth noting that there is no obvious correlation between the jet power and active age in \cref{fig:fitted_sample_energetics}, suggesting we can reasonably ignore higher-order interactions between these parameters in our work; i.e., we can assume the jet power and active age are only weakly correlated. Regardless, in the following section, we will investigate their distributions in $(Q, t_{\mathrm{on}})$-space, rather than the marginal PDFs, to ensure we constrain these parameters independently of each other.

\section{Simulating the observed energetics}
\label{sec:mock_remnant_populations}

The jet power and lifetime distributions fitted for our remnant sample in \cref{sec:observed_jet_age_dist} will not represent the underlying population statistics due to the sample selection biases. We outline a process to resolve this problem by comparing the energetics of a mock population generated with known underlying distributions -- with sample selection biases applied -- to those fitted for our observed remnant sample. First, we generate mock remnant catalogues for an assumed jet power and lifetime distribution, and model how the selection bias modifies the observed energetics (\cref{sec:modelling_the_selection_bias}). Second, we must control for the confounding variables (i.e., redshift, halo mass, equipartition and injection index) so that the predictions made for a given jet power and lifetime distribution can be meaningfully compared to our actual sample (\cref{sec:controlling}).

\subsection{Modelling the sample selection bias}
\label{sec:modelling_the_selection_bias}
Mock catalogues of remnant radio galaxies are created by weighting the probability of the parameter space investigated in the RAiSE simulation. To model the effect selection bias has on the observed energetics, two separate weighting schemes are required. First, the simulation needs to be weighted by an underlying jet power and active age distribution (\cref{sec:input_Q_ton}); we refer to this weighting as $W_{\mathrm{jet\text{-}age}}$, which controls the assumed shape of the underlying distributions we will later constrain. Second, to compute the selection bias, $W_{\mathrm{selection}}$, a binary weighting is assigned to each synthetic source depending on whether or not the source would pass the sample selection criteria (\cref{sec:applying_the_selection_criteria}).

\subsubsection{The input jet power and lifetime distributions}
\label{sec:input_Q_ton}
We begin with the assumption that a simple power-law model is sufficient to describe the distributions of the radio-loud AGN jet powers and lifetimes. Observational constraints on the AGN radio luminosity function (RLF) show that the shape of the RLF is a broken power law \citep[e.g., see][]{2019A&A...622A..17S,2021PASA...38...41F}. Considering that, to first order, the radio luminosity is correlated with jet power (noting that the age and environment also have a confounding impact, e.g., \citealt{2013ApJ...767...12G,2013MNRAS.430..174H}), and that our sample preferentially selects high-powered remnants (see \cref{fig:fitted_sample_energetics}), we assume our assumption holds for the jet power distribution. Meanwhile, simulations of feedback-regulated black hole accretion show that the Eddington-scaled mass accretion rate, $\dot{m} = \dot{M}_{\mathrm{BH}}/M_{\mathrm{Edd}}$, follows a power spectrum consistent with pink noise \citep[e.g., see][]{2011ApJ...737...26N,2017MNRAS.466..677G}. These simulations predict that the jets triggered at the black hole are described by a lifetime function consistent with a $p(t_{\mathrm{on}}) \propto t_{\mathrm{on}}^{-1}$ power-law slope\footnote{This function describes the likelihood of a jet being triggered with an active age $t_{\mathrm{on}}$, not the relative probability of observing such a source.}. Based on our discussion in \cref{sec:observed_jet_age_dist}, we assume that the two distributions (jet power and active age) are seeded independently; that is, that no intrinsic correlation exists between these two quantities. Future work based on a larger sample could investigate more complex model assumptions including correlations with the host galaxy mass (see \cref{sec:ch4_conclusion}). 

Based on these arguments, we model the underlying jet power and lifetime distributions as the following probability density functions (PDFs):
\begin{subequations}
\begin{gather}
    p(Q) \,dQ = \left({Q}/{Q_0}\right)^{a} \,dQ ,\label{eqn:Q}\\
    p(t_{\mathrm{on}}) \,dt_{\mathrm{on}} = \left({t_{\mathrm{on}}}/{t_0}\right)^{b} \,dt_{\mathrm{on}},\label{eqn:ton}
\end{gather}
\end{subequations}
where $Q_0$ and $t_0$ are arbitrary constants that ensure the total probability is unity between the lower and upper bounds (see \cref{tab:grid_properties}). The coefficients $a$ and $b$ give the power-law indices for the jet power and lifetime distributions of each jet outburst (i.e., not considering the likelihood of observing). 

However, as the input parameter space for the RAiSE simulation is uniformly gridded in $\log Q$ and $\log t_{\mathrm{on}}$, weighting the simulation using the above expressions would fail to account for the non-uniform grid spacing when transformed to linear space. This is important considering that, for example, the grid bin at $\log (t_{\mathrm{on}}/\text{yrs})=7.5$ covers is an order of magnitude greater range of ages than that at $\log (t_{\mathrm{on}}/\text{yrs})=6.5$. As such, the above expressions are adapted by the bin correction as follows:
\begin{subequations}
\begin{gather}
    p(\log Q) \,d\log Q = \left(Q/Q_0\right)^{a + 1} Q_0 \ln 10 \,d\log Q ,\label{eqn:jet_corr}\\
    p(\log t_{\mathrm{on}}) \,d\log t_{\mathrm{on}} = \left(t_{\mathrm{on}}/t_0\right)^{b + 1} t_0 \ln 10 \,d\log t_{\mathrm{on}} .
\label{eqn:age_corr}
\end{gather}
\end{subequations}
We will not provide results in the form of these equations -- these are only necessary to weight the simulation outputs to the correct, temporally uniform sampling assumed in \cref{eqn:Q,eqn:ton}.  

The remnant radio lobes modelled in this work are observed exclusively after the jet has switched off (by definition)\footnote{Restarted radio sources included in our sample may have an active jet confined to the host galaxy -- in this case, our methods are explicitly modelling the outburst associated with the remnant lobe.}. The active age is correctly weighted for uniform sampling in the linearly-spaced active age parameter (\cref{eqn:ton}), but we have not considered if this holds for the remnant phase, parameterised through the remnant ratio $R_{\mathrm{rem}} = t_{\mathrm{rem}}/(t_{\mathrm{on}} + t_{\mathrm{rem}})$. We find the following relationship between a uniform increment in the linearly-spaced remnant ratio (see \cref{tab:grid_properties}) and a uniform increment in off-time:
\begin{equation}
    dt_{\mathrm{rem}} = \frac{(t_{\mathrm{on}} + t_{\mathrm{rem}})^2}{t_{\mathrm{on}}}\;dR_{\mathrm{rem}}.
\label{eqn:R_corr}
\end{equation}
This equation implies that simulation outputs for a given remnant ratio must be weighted by the factor $(t_{\mathrm{on}} + t_{\mathrm{rem}})^2/t_{\mathrm{on}}$ to ensure uniform sampling in the off-time. We apply this correction to our simulation outputs.

The weightings in \cref{eqn:jet_corr,eqn:age_corr,eqn:R_corr} specify the weighting required both to encode the underlying jet power and lifetime functions, and to convert from the sampling in our simulation grid to a temporal uniform sampling. The function $W_{\mathrm{jet\text{-}age}}$ represents the weights calculated in this manner for each point in the $k=6$ parameter space, enabling us to generate mock populations, albeit without selection effects applied as yet.

\subsubsection{Applying the selection criteria}
\label{sec:applying_the_selection_criteria}

The selection biases inherent to the observed sample (\cref{sec:parent_sample}) are replicated for each mock source in the RAiSE simulation grid by assessing their synthetic surface brightness images against our sample selection criteria. Three key criteria merit consideration:~(1) the minimum integrated flux density;~(2) the minimum surface brightness; and~(3) the minimum angular size measured along the radio lobes. The modelling of each selection criteria is discussed below, and is intended to replicate the steps outlined in \cref{sec:remnant_sample} for the observed sample.

To construct the parent catalogue of radio sources, the GLEAM catalogue was used to select radio sources brighter than $0.5$\,Jy at 154\,MHz (i.e., $S_{154}\geqslant0.5\,$Jy). For each synthetic radio source, the 154\,MHz integrated flux density is calculated based on the forward-modelled radio luminosity and the given mock redshift. As such, we assume that the synthetic sources fainter than this threshold would not have made it into our sample. Considering that the GLEAM catalogue contains only the components with a signal to noise larger than $5\sigma$, where $\sigma$ is the GLEAM local rms in Jy/beam, this sets the condition for the minimum surface brightness criteria. To do this, the synthetic 154\,MHz synchrotron emission maps are convolved to match the GLEAM resolution at 154\,MHz. Using these surface brightness maps, we assume that synthetic sources with a peak flux density fainter than the $5\sigma$ GLEAM sensitivity would not have been detected, and thus will be missed by our selection. Finally, our sample contains radio sources with a projected angular size greater than $\theta=4'$ in RACS-low. To model this selection criteria, the synthetic 887\,MHz synchrotron emission maps are reprojected for a given mock redshift, and convolved to match the RACS-low resolution at 887\,MHz -- we assume all the observed radio sources lie in the plane of the sky. Considering only the regions of the lobes brighter than $750\,\mu$Jy/beam ($\approx3\sigma$), we use the surface brightness maps to discard sources whose largest angular size falls below our sample threshold. To model the bias due to these sample selection criteria, we construct a binary selection grid as follows:
\begin{equation}
W_{\mathrm{selection}} = \bigg\{ \!\begin{array}{ll}
    1,&\text{selected} \\
    0,&\text{not selected}
  \end{array},
\end{equation}
where the synthetic source is weighted either as unity (the source passes the selection criteria), or zero (the source fails one or more of the selection criteria). 

We note, the framework above only models the selection due to the parent sample selection criteria (i.e., \cref{sec:parent_sample}), and does not consider the potential selection biases due to the remnant selection methods we have used. For this work, we assume that as soon as a source switches off it will show observable signatures consistent with a remnant radio galaxy; in other words, the remnant radio galaxies forward modelled by our mock catalogues are ones that we should be able to classify. This assumption may not hold for extremely young remnants, for example those in which ${t_\mathrm{rem}}$ is smaller than the light travel time along the jet, ${t_\mathrm{light\text{-}travel}}$\footnote{For a distance of $100$\,kpc, information travelling at a speed $0.3c$ would take $\sim 1$\,Myr to reach the hotspot from the SMBH.}, however, we do not expect this to be a problem for the large off-times probed by our sample.

Following this approach, the selection of each source is evaluated for the entire RAiSE simulation. Importantly, by evaluating $W_{\mathrm{jet\text{-}age}}$ and $W_{\mathrm{selection}}$, predictions for the observed jet power and lifetime distributions can be made for any underlying assumptions and sample selection criteria. To make predictions for the observed energetics, the RAiSE simulation grid is sampled using the element-wise multiplication of the two weighting schemes: i.e., $W_{\mathrm{jet\text{-}age}} \circ W_{\mathrm{selection}}$. 

This approach is demonstrated in \cref{fig:selected_energetics_initial_prior} assuming a mock population of $z=0.1$ remnant galaxies for jet power and active age distributions of $p(Q)\propto Q^{-1.2}$ and $p(t_{\mathrm{on}})\propto t_{\mathrm{on}}^{-0.5}$ respectively. The sample selection bias against lower-powered and shorter-lived sources is clearly seen: these sources are fainter and have a smaller physical sizes, and are thus precisely the objects that would be preferentially deselected from our sample. Higher-powered jets produce stronger lobe magnetic fields, in turn resulting in shorter fading timescales during the remnant phase \citep[e.g., ][]{2018MNRAS.474.3361T,2018MNRAS.476.2522T,2022MNRAS.514.3466Q}; this effect is likely responsible for the observed flattening towards $Q=10^{40}$\,W. Similarly, the downturn observed towards the longest-lived sources is likely a combination of two factors: the expansion of the source driving down the lobe surface brightness \citep[e.g.,][]{2018MNRAS.476.2522T,2020MNRAS.496.1706S}, and the decrease in the integrated radio luminosity due to Rayleigh-Taylor mixing of the lobe plasma with the environment (e.g., see Fig.~3 of \citealt{2015ApJ...806...59T}). Importantly, these results show us sample selection biases can greatly transform the observed nature of the true jet power and active age distributions.

Our approach so far, however, has neglected the impact the remaining sample properties will have on these results. This is discussed in the following section.

\subsection{Controlling for the global properties of the sample}
\label{sec:controlling}

The range of plausibly-selected jet power and lifetime distributions are strongly confounded by the redshift, in addition to the halo mass, equipartition factor, and injection index. The selection biases computed for our mock catalogues must therefore take into account how these parameters are distributed for our sample. To do this, we initially apply a zeroth-order correction: assume that their underlying distributions are the same as what we observe (\cref{sec:initial_priors}). However, after applying the sample selection biases the PDFs for each parameter will likely differ from their input distributions; we therefore apply higher-order corrections by iteratively perturbing the underlying distributions of these parameters until -- upon applying the selection effects -- they match the observed sample (\cref{sec:correcting_for_selection_bias}).

\subsubsection{Weighting the initial priors}
\label{sec:initial_priors}

We use the observed statistics in each parameter to weight their corresponding mock distributions; e.g., the observed redshift distribution acts as a prior to weight the redshift distribution of the synthetic sources, as we now describe in detail. The observed redshifts are binned to to the closest matching mock redshift, along a uniformly spaced axis from $(0,0.2]$ using a resolution of $dz=0.005$\footnote{The RAiSE simulations run at $z=0.1$ are scaled to these mock redshifts for computational efficiency, as discussed in \cref{sec:input_parameter_space} for our remnant sample fitting.}. This assignment of the observed redshifts is repeated using 1000 Monte Carlo (MC) realisations of the each redshift within its measurement uncertainties. The resulting distribution is normalised to unity; this normalised PDF for the redshift is taken as the prior to weight our simulation outputs. We note, empty bins are assigned a weighting of zero, meaning there is a null probability of finding a source with that particular parameter value. 

The observed distributions of the other confounding variables are built up using MC statistics as for the redshift, with the simulation grid for each parameter specified in \cref{tab:grid_properties}. The uncertainties on these confounding variables are taken as the fitted error when modelling the evolutionary histories of each remnant in \cref{sec:maxlikelihood_gridsearch}; i.e., unlike the redshift, we do not have robust measurements for these parameters before the RAiSE-based inversion. In this way, we have characterised the observed distributions of each confounding variable and converted these to prior PDFs for each axis of the RAiSE simulation grid. 

\begin{figure*}
    \centering
    \includegraphics{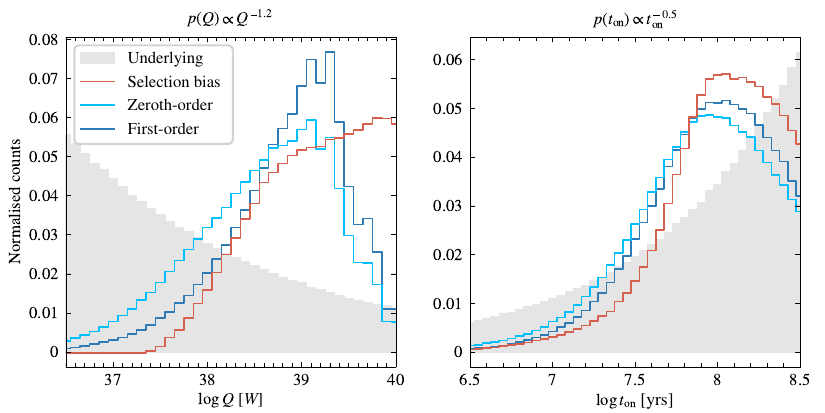}
    \caption{The underlying distributions (grey bars) in the jet power (left panel) and active age (right panel) with $p(Q)\propto Q^{-1.2}$ and $p(t_{\mathrm{on}})\propto t_{\mathrm{on}}^{-0.5}$ respectively. This mock catalogue is simulated at $z=0.1$, $\log (H/\text{M}_\odot)=14$, $\log q=-1.7$ and $s=2.2$. We filter this catalogue through our sample selection criteria (orange; \cref{sec:parent_sample}). The prior weightings for the redshift, halo mass, and other confounding variables, are derived from the sample distributions as either a zeroth-order correction (light blue; \cref{sec:initial_priors}) or are updated iteratively to match the sample distributions, acting as a first-order correction (dark blue; \cref{sec:correcting_for_selection_bias}).}
    \label{fig:selected_energetics_initial_prior}
\end{figure*}

The zeroth-order correction to our mock population is simply to weight each axis of the RAiSE simulation grid by the observed distributions in these confounding variables. We apply these prior PDFs to the example population considered in the previous section (shown in \cref{fig:selected_energetics_initial_prior}), demonstrating the effect of considering the remaining properties of our sample. The most notable difference is the turn-off towards the highest-powered jets; a feature that is also observed for our remnant sample (\cref{fig:fitted_sample_energetics}). This is likely due to some higher-powered jets not forming lobes in the lowest-density environments \citep[i.e., balistic expansion;][]{1991MNRAS.250..581F,2023Galax..11...87T}.

\subsubsection{Controlling for confounding variables}
\label{sec:correcting_for_selection_bias}
\begin{figure*}
    \centering
    \includegraphics{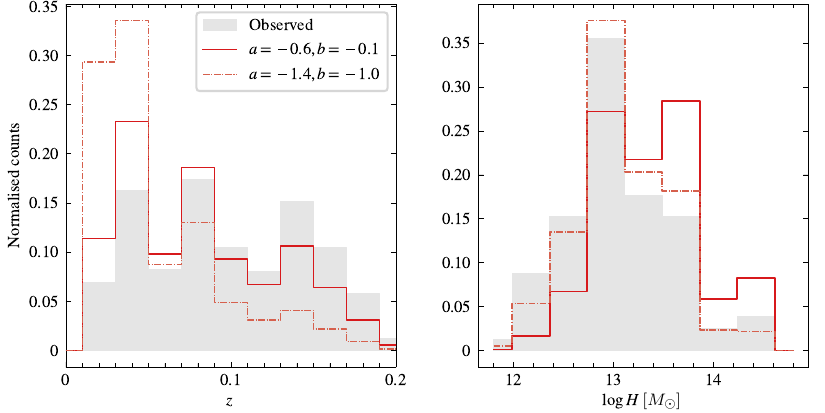}
    \caption{The observed redshift (left panel) and halo mass (right panel) distributions predicted for two separate mock catalogues (solid red and dot-dashed orange steps) created for different jet power and active age distributions (see exponents in figure label). Mock catalogues are filtered by the selection criteria of our sample, and are weighted by the distributions observed for the redshift and the confounding variables (grey bars). The redshift bins plotted here are larger than those of the mock simulation grid to aid this visualisation.}
    \label{fig:uncorrected_ZH_dist}
\end{figure*}

We have so far applied a zeroth-order correction to our confounding variables by assuming their underlying distributions match those observed for our sample. However, this is unlikely to be true as, like for the jet power and active age, the distributions in these parameters will be moderated by the sample selection effects. We demonstrate this problem as follows: we generate two mock catalogues with different underlying energetics, weight their confounding variable PDFs following the method outlined in the previous section, and finally apply the selection criteria of our sample. The mock redshift and halo mass PDFs -- resulting after applying the selection biases -- are compared to the observed distributions in \cref{fig:uncorrected_ZH_dist}, showing the selection bias does indeed moderate the shape of the observed distributions. The output redshift and halo mass PDFs do not match their input distributions, implying that the observed statistics of our remnant sample must have been selected from a currently-unknown underlying distribution. 

The manner by which the outputs differ to the input is consistent with our expectation. The output redshift PDF is skewed towards lower-redshifts: for a fixed jet power and lifetime distribution, the highest redshift bin will have the narrowest range of plausibly-selected evolutionary histories. As for the halo mass distribution, the bias is evident in lower mass environments where high-powered jets are less likely to form lobes. In \cref{fig:uncorrected_ZH_dist}, it is also apparent that the selection bias acting on the redshift and halo mass is dependent upon the underlying energetics (red versus orange lines). The model with more low-powered and short-lived sources ($a = -1.4$ and $b=-1.0$) tends to be deselected at higher redshifts (fewer bright sources) and in higher mass haloes (fewer jets meeting the angular size cut). 
Overall, these findings imply that the underlying distributions for the confounding variables must have higher-order corrections applied such that the resulting survey-selected PDFs match those of our observed sample. 

The higher-order corrections are derived by iteratively applying corrections to the confounding variable PDFs as in \cref{sec:initial_priors}. We apply the zeroth-order correction as before, then, for each confounding variable, we compare the error in each bin (\cref{tab:grid_properties}) between the observed and mock PDFs for a proposed jet power and lifetime distribution; i.e., $N_{\mathrm{observed}}/N_{\mathrm{mock}}$. This ratio quantifies how significant the selection bias is at each bin, which varies somewhat smoothly across the parameter space. The prior PDFs are updated by the inverse of this ratio to up-weight values that were under-represented after the previous correction. The mock catalogues are re-sampled with the updated prior PDFs for each of the confounded variables. This process can be repeated over several iterations, however, we find that a single round of (first-order) corrections is sufficient to bring the mock distributions in agreement with our remnant sample. 

Finally, with these corrections, we are able to produce mock catalogues that are correctly weighted for their redshift and other confounding-variable distributions. The predictions made by these catalogues, e.g., see \cref{fig:selected_energetics_initial_prior}, can now compared to the observed jet power and active age distribution of our sample, in order to constrain their underlying distributions. This is explored in the following section.

\section{Results and discussion}
\label{sec:results}

We now can separate the impact of selection bias from our observed jet power and lifetime distributions using the method outlined in \cref{sec:mock_remnant_populations}. This allows us to infer their underlying distributions (\cref{sec:Constraining the seed distributions}) and compare these findings with previous work in the literature (\cref{sec:literature_consistency}). In \cref{sec:jet_triggering}, we discuss our results in the context of jet triggering and fuelling mechanisms, and then we suggest techniques to apply our approach to the active radio source population in \cref{sec:Application to active radio galaxy populations}.

\subsection{Constraining the underlying distributions}
\label{sec:Constraining the seed distributions}
Following the methodology described in \cref{sec:mock_remnant_populations}, we generate mock catalogs in a two-dimensional grid of proposed jet power and lifetime distributions, where each cell corresponds to the power-law indices $(a,b)$. The power-law index for the jet power distribution is binned into discrete intervals in $-2.6 \leqslant a \leqslant -0.6$ using a precision of $\delta a=0.1$. Similarly, the power-law index for the lifetime distribution is partitioned into intervals in $-2 \leqslant b \leqslant -0.5$ with a precision of $\delta b=0.05$. 

\begin{figure*}
    \centering
    \includegraphics{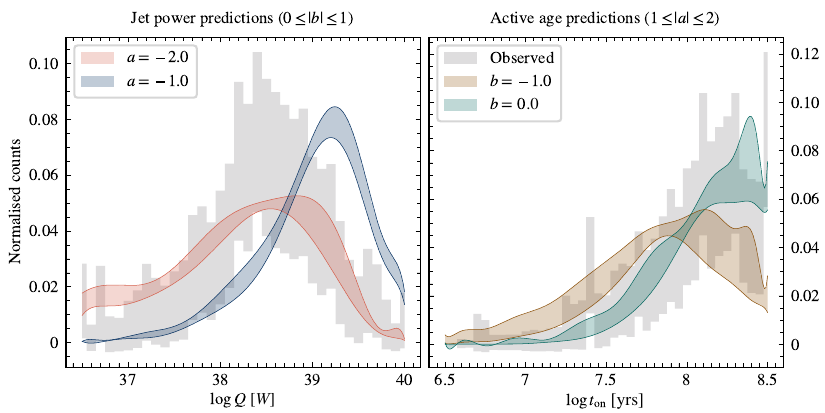}
    \caption{Predictions for the observed distributions in jet power (left panel) and active age (right panel) made by various mock populations. The left panel shows predictions made by a steep ($a=-2.0$, red) and shallower ($a=-1.0$, blue) jet power model, evaluated for all age distributions. The right panel shows predictions made by a steep ($b=-1.0$, brown) and shallower ($b=0$, teal) ageing model. evaluated for all jet power models with $1\leqslant |a| \leqslant 2$. The mock distributions are smoothed using a $0.3$\,dex kernel.}
    \label{fig:jetage_diagnostic}
\end{figure*}

We first investigate whether our mock catalogues (for the proposed of underlying distributions) exhibit discernible differences when influenced by the selection biases. We compare the predicted jet power distribution between two scenarios: a steep jet power distribution with $a=-2.0$ (many low-powered jets) and a shallow distribution with $a=-1.0$ (fewer low-powered jets), and consider all age models within $-2 \leqslant b \leqslant -1$. Similarly, we compare the predicted lifetime distribution between the following two scenarios: a steep distribution with $b=-2.0$ (many short-lived jets) and a shallower distribution with $b=-1.0$ (fewer short-lived jets), and consider all jet power models within $-2\leqslant a\leqslant-1$. \cref{fig:jetage_diagnostic} presents our findings, which are compared with the observed sample statistics from \cref{fig:fitted_sample_energetics}. Encouragingly, despite the stringent selection criteria, the shape of the predicted distributions (after selection biases are applied) are measurably different: those corresponding to shallower underlying distributions exhibit a turn-over at approximately one order of magnitude greater than the steeper underlying distribution. 

\cref{fig:jetage_diagnostic} further illustrates the intertwined relationship between the underlying jet power and active age distributions, and their selection-biased counterparts. For a given jet power model, the shape of the observed jet power distribution is somewhat influenced by the underlying jet lifetime model, and vice versa. This indicates that while the observed distribution primarily constrains its corresponding underlying distribution (i.e., observed jet powers constrain the jet power model and observed active ages constrain the lifetime model), there should in principle be weaker constraints stemming from the complementary distribution (i.e., observed active ages aiding in constraining the jet power model, and vice versa).

\begin{figure*}
    \centering
    \includegraphics{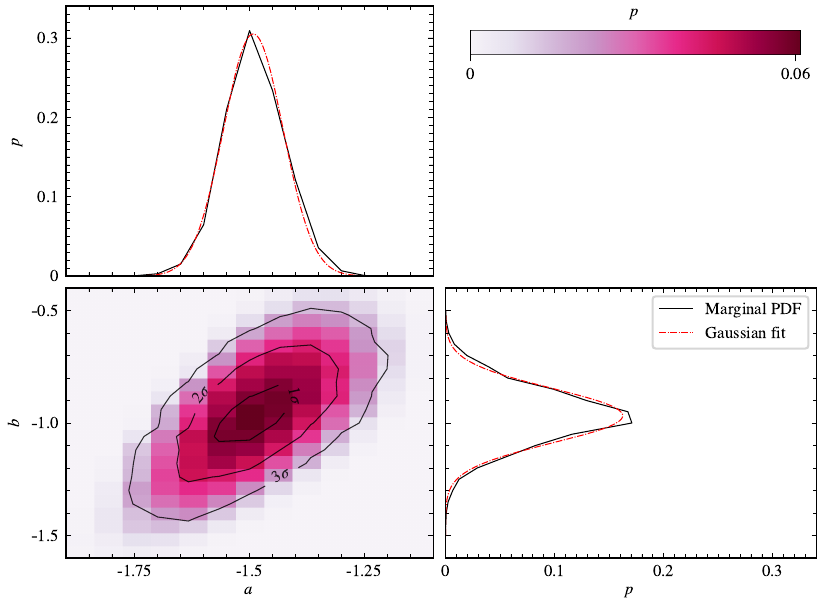}
    \caption{The relative fitting probabilities evaluated for each jet power and lifetime model. The corresponding probability density function (PDF) mapped in $( a,b )$-space is shown in the bottom left panel. A sequential colour-map is used to indicate the probability for each model, and is shown using a logarithmic stretch. Contours denote the $1\sigma$, $2\sigma$ and $3\sigma$ confidence intervals. The adjacent panels demonstrate the PDF marginalised for $a$ (upper left panel), and $b$ (lower right panel), as shown by the solid black curves. The red dot-dashed curves show the Gaussian approximations of the marginal PDFs. }
    \label{fig:powerlaw_indices}
\end{figure*}

To constrain the evolutionary jet model, the mock jet power and active age distributions are binned in $( Q, t_{\mathrm{on}})$-space to match the binning of the observed distributions (see \cref{fig:fitted_sample_energetics}). The likelihood function is evaluated by considering the mock and observed constraints at each point in this parameter space, from which we calculate the AIC for two free parameters. This calculation assumes that each data point (pixel) offers an independent measurement on the number of sources in that particular $( Q, t_{\mathrm{on}})$ bin, which is not strictly true considering that there is a correlation between the number of sources in neighbouring bins. We note however that the uncertainty on the observed $( Q, t_{\mathrm{on}})$ distribution is estimated via Monte Carlo simulation (see \cref{sec:observed_jet_age_dist}), and encodes the effect of this correlation. We select the optimal jet model as that which minimises the AIC, and we compute the relative probability of each model as $p_i = e^{(\mathrm{AIC}_i - \mathrm{AIC}_\mathrm{best})/2}$. The 2D probability density function (PDF) is mapped across $( a,b )$-space (\cref{fig:powerlaw_indices}); the PDF shows some covariance, implying that the jet power and lifetime distribution are not fitted completely independent of each other.

The parameter estimates are quantified using their 1D marginal PDFs, shown in \cref{fig:powerlaw_indices}. We approximate the marginal PDFs using a Gaussian function to characterise the most probable value and the $1\sigma$ standard deviation; this reduces the need to sample the parameter space more finely at large computational expense. As such, we constrain the power-law slopes of the underlying jet power and lifetime functions as follows:
\begin{subequations}
\begin{gather}
    p(Q)\propto Q^{-1.49 \pm 0.07} ,\\
    p(t_{\mathrm{on}})\propto t_{\mathrm{on}}^{-0.97 \pm 0.12} .
\end{gather}
\end{subequations}

\begin{figure*}
    \centering
    \includegraphics{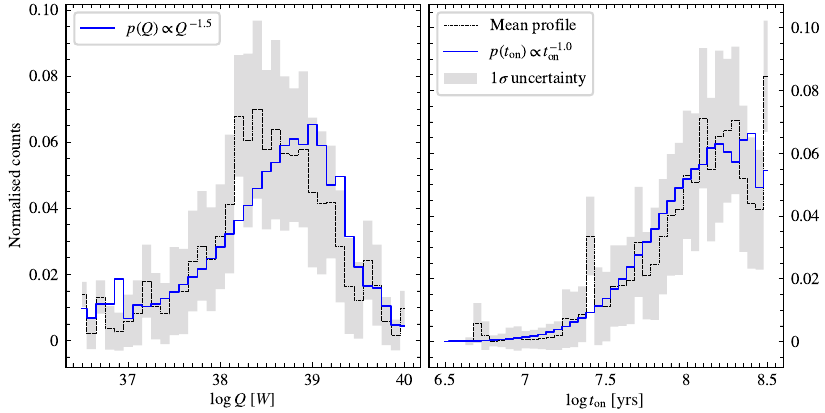}
    \caption{Same as \cref{fig:fitted_sample_energetics}, but overlaying the predictions made by our best-fit jet power and lifetime model; i.e., $p(Q)\propto Q^{-1.5}$ and $p(t_{\mathrm{on}})\propto t_{\mathrm{on}}^{-1.0}$.}
    \label{fig:underlying_energetics}
\end{figure*}

The resulting survey-selected jet power and lifetime distributions are compared to the observed distributions in \cref{fig:underlying_energetics}. We can see that the shape of the mock active age distribution agrees with observation; the mock distribution follows the same initial rise in number counts as the observed distribution towards longer-lived sources, and exhibits a similar turnover at $t_{\mathrm{on}} = 10^{8.25}$\,yrs. Within uncertainties, the general shape of the mock jet power distribution also appears consistent with observation; the number density of sources at the higher and lower-powered ends are in agreement with each other, and the turn overs in each distributions are comparable. However, several potential inconsistencies should be pointed out. Firstly, the turnover in the mock jet power distribution ($Q=10^{39}$\,W) occurs roughly an order of magnitude higher than that in the observed distribution ($Q\approx 10^{38.2}$\,W). Secondly, the rate at which the mock distribution falls off towards low-powered sources is not as steep as what appears to be the case for our observed sample. In fact, by manually inspecting the predictions made our mock catalogues (but see \cref{fig:jetage_diagnostic}), we find that no model is able to simultaneously replicate each of these features. Flatter values of $a$ give the correct fall-off towards low $Q$, but turn over at much higher jet powers. Steeper values of $a$ predict a closer turn over to what is observed, but over predict the number density of low-powered objects. 

We speculate that one potential explanation could be that the true jet power function flattens below a certain jet power, $Q_*$, which would be expected if the broken power-law nature of the AGN radio luminosity function is driven directly by shape of the jet power function. Such a model should in principle provide a solution to the two inconsistencies outlined above; by flatting below some $Q_*$, the expected fall-off towards low-powered sources should be sharper than what is currently shown by our mock catalogue in \cref{fig:fitted_sample_energetics}. Considering our relatively low sample size $(N=79)$, we have chosen not to fit a broken power-law jet power model for this work. Nevertheless, under the assumption that the jet power function is indeed more appropriately described by a broken power law, our results here show a promising future application for radio galaxy populations revealed by next-generation radio surveys.

\subsection{Comparison with the literature}
\label{sec:literature_consistency}

Forward modelling has recently been employed in two published works to constrain the bulk properties of radio-loud AGNs based on the distributions in their observed attributes.
\citet{2019A&A...622A..12H} used the dynamical model of \citet{2018MNRAS.475.2768H} to apply forward modelling to radio-loud AGNs unveiled by LoTSS DR1. As part of their analysis, they tested the validity of two different lifetime models in which the observed active ages were distributed uniformly in either linear age ($b=0$ in this work), or in logarithmic age ($b=-1$ in this work). Their jet power function was assumed to be $p(Q)\propto Q^{-1}$. To examine the validity of their lifetime models, the linear size distributions were forward modelled for each model assumption, and compared to the observed size distributions for three radio luminosity bins. Their main finding (see their section~4.2) is that neither lifetime function was able to simultaneously explain the observed size distributions across all luminosity bins. Their linear age model gave excellent agreement with the size distributions in the brightest two bins, however vastly under-predicted the fraction of small ($<100\,$kpc) radio sources in the faintest bin. On the other hand, at the highest two luminosity bins, their log-uniform lifetime function vastly over-predicted the compact fraction, and simultaneously under-predicted the observed size distributions. 

\citet{2020MNRAS.496.1706S} employed RAiSE forward-modelling to constrain the jet power and lifetime functions for a LOFAR-selected sample of radio galaxies. For a given jet power and lifetime function, they forward-modelled radio galaxy populations, applied sample selection biases, and compared their predicted angular size and flux density distributions to those observed for their sample. In particular, they tested power-law and constant-age models, and found that both could plausibly explain the observed properties of their active radio sources. To break down the degeneracy between these two lifetime functions, they compared the mock remnant and restarted fractions with observation, and found that power-law lifetime function offered much better agreement. In particular, they found that a model with $p(Q)\propto Q^{-1}$ and $p(t_{\mathrm{on}}) \propto t_{\mathrm{on}}^{-1}$ offered excellent agreement with their observed sample, with $p(t_{\mathrm{on}}) \propto t_{\mathrm{on}}^{-0.5}$ also offering good agreement. 

We now examine the consistency between our results and those of the two aforementioned studies, but note that differences are expected due to differences in the methodologies we have adopted and the radio galaxy samples we have used.
We constrain the same optimal lifetime model as \citet{2020MNRAS.496.1706S}, i.e., $p(t_{\mathrm{on}}) \propto t_{\mathrm{on}}^{-1}$, though our results show their $p(t_{\mathrm{on}}) \propto t_{\mathrm{on}}^{-0.5}$ model is incompatible with the observed active age distribution. These findings are also tentatively in agreement with \citet{2019A&A...622A..12H}. 
Considering that we have directly measured the full lifetime of the jets in these sources, we would expect the fitted active ages to occupy a single value (within fitting uncertainties) if their underlying lifetimes were that of a constant age. 
On the other hand, both \citet{2019A&A...622A..12H} and \citet{2020MNRAS.496.1706S} find considerably flatter jet power models, which may be a result of the sample selection. In \cref{sec:Constraining the seed distributions}, we speculated that the true jet power function may be more appropriately described by a broken power law, such that a flattening would occur below some critical value. The jet powers fitted by \citet{2020MNRAS.496.1706S} for their sample show a peak in their distribution at $Q \approx 10^{37.7}$\,W (see their fig. 3), almost an entire order of magnitude lower than our sample. We therefore speculate that the flatter jet power distribution constrained in their work could simply be a consequence of their sample probing a lower-powered portion of the jet power function {(discussed further in \cref{sec:jet_triggering})}.

\subsection{Jet triggering and fuelling mechanisms}
\label{sec:jet_triggering}

A key result of this work is the confirmation that the jet lifetime function is consistent with $p(t_{\mathrm{on}}) \propto t_{\mathrm{on}}^{-1}$, implying that most radio sources are short-lived. The active age distribution fitted for our remnant sample offers direct evidence against constant age models ($>$$5\sigma$ confidence level), and is in support of the conclusions of \citet{2020MNRAS.496.1706S}. 
This finding begs the question: what jet triggering mechanisms lead to more short-duration outbursts?
Simulations of feedback-regulated black hole accretion consistently predict a time-varying mass accretion rate. The associated power spectrum is consistent with pink noise, corresponding to a lifetime function $p(t_{\mathrm{on}}) \propto t_{\mathrm{on}}^{-1}$ \citep[e.g., see][]{2011ApJ...737...26N,2017MNRAS.466..677G}. This agreement with our jet lifetime function, and earlier work by \citet{2020MNRAS.496.1706S}, strongly suggests the jet active age distribution is moderated by feedback-regulated accretion.

{The high jet powers modelled for our remnant sample may, however, be difficult to obtain with an advection dominated accretion flow \citep[ADAF; e.g., equation 7 of][]{2001Meier}; i.e., $Q \lesssim 10^{38.3}$\,W for a massive $M_\bullet = 10^9$\,M$_\odot$ black hole with an Eddington-scaled accretion rate of $\dot{m} < 0.03$ \citep[e.g.,][]{Park_2001}. This suggests that the majority of our remnant sample was fuelled (in their active state) by either a thin or slim disk accretion flow; or, alternatively, their host black holes are atypically massive ($M_\bullet > 10^9$\,M$_\odot$; e.g., cf. \citealt{Schulze+2010}) and accreting at close to $\dot{m} \approx 0.03$. The radiatively efficient thin disk is less effective at producing jets than an ADAF, in principle, requiring super-Eddington accretion rates to match even the lowest jet powers in our sample \citep[cf. equation 5 of][]{2001Meier}. The slim disk accretion flow, in contrast, can generate jet powers across the full range probed by our sample; slim disks are stable for slightly super-Eddington accretion rate, $\dot{m} \gtrsim 1$ \citep[e.g.,][]{1988ApJ...332..646A, Park_2001}. The high accretion rates necessitated by this accretion flow (under this hypothesis) suggest the jet triggering mechanism relevant for much of our sample may be cold-mode accretion: streams of cold gas from galaxy interactions or minor mergers. The inconsistent jet power functions fitted in this work and that of \citet[][discussed in \cref{sec:literature_consistency}]{2020MNRAS.496.1706S} may therefore be explained by different jet triggering mechanisms; i.e., cold-mode accretion versus feedback-regulated accretion, respectively.}

We note that numerous authors have suggested that different jet triggering mechanisms may mediate the lifetime functions for different radio source populations \citep[e.g.,][]{2013MNRAS.429.1827P, 2015MNRAS.452..774K, 2018MNRAS.474.3615M}. {Compared to other authors, our remnant sample probes a population with particularly high jet kinetic powers, approximately an order of magnitude higher than that of \citet{2020MNRAS.496.1706S}. Intriguingly, the pink noise spectrum indicative of feedback-regulated accretion is encoded over the several orders of magnitude in AGN jet power probed by these samples -- though of course this may instead suggest cold-mode accretion follows the same spectrum.} Regardless, neither sample considers significant populations of lower-powered \citet{1974MNRAS.167P..31F} Type-I sources ($Q \lesssim 10^{37}$\,W) so our conclusions may be limited to the higher-powered Type-II class.

\subsection{Application to active radio galaxy populations}
\label{sec:Application to active radio galaxy populations}
While the approach taken for this work focused solely on remnant radio galaxies, our methods for simulating mock radio source populations can be generalised to the wider radio-loud AGN population; specifically to active radio galaxies {(cf. comparable studies by \citealt{2015ApJ...806...59T} and \citealt{2019A&A...622A..12H})}, which make up $\sim80\%$ of the observed radio source population \citep[e.g., see][]{2020A&A...638A..34J}. As mentioned in \cref{sec:ch4_introduction}, the age measured for an active radio galaxy represents the age at which the source is observed, $\tau$, not the true duration of its active phase {(i.e., $t_\text{on}$)}. This pushed \citet{2015ApJ...806...59T} to assume that, on average, active radio sources will be observed halfway through their active lifetime, i.e., $\tau = t_{\mathrm{on}}/2$. {However, an alternative approach to this problem would be to simulate an observed $\tau$ distribution, filtered by the selection criteria outlined in \cref{sec:mock_remnant_populations}, based on each proposed jet lifetime function, $p(t_\text{on})$; e.g., $p(\tau)d\tau \propto -\ln(\tau) d\tau$ for $p(t_\text{on}) \propto t_\text{on}^{-1}$, ignoring selection biases\footnote{The age distribution for active sources is the integral of the lifetime function (i.e., the sum over all sources that are active at a given age $\tau$) as their evolution is independent of the active age for $\tau \leqslant t_\text{on}$}.} The simulated, selection-biased age distribution{, $p(\tau)$,} can be directly compared to the source ages, $\tau$, fitted for active radio galaxies. Of course, a major caveat with this method is that a large observational sample (e.g., $\sim$$1000$ sources) would be required to confidently average over the random noise resulting from sampling each active source at some unknown point in its active phase -- the statistical ensemble resulting from numerous objects is needed to inform the active age in this case.

\section{Conclusion}
\label{sec:ch4_conclusion}

We have presented an approach to constrain the jet power and active age functions for active galactic nuclei (AGN) jet outbursts by decoding observations of remnant radio galaxies. These objects uniquely measure the full duration of the active phase, thus placing firm constraints on the jet lifetime function. We compiled a sample of remnant radio galaxies using large-sky multi-frequency radio data with a fixed selection criteria, i.e., $S_{154}\geqslant 0.5$\,Jy and $\theta\geqslant4'$, and identified remnant lobes using variety of literature-established methods (\cref{sec:remnant_sample}). The energetics of each source were constrained using the RAiSE dynamical model; simulated radio source attributes were compared with observation through a parameter inversion, which allowed fitting of, amongst other parameters, their jet power and active age (\cref{sec:sample_energetics}). However, due to sample selection biases, the fitted jet power and active age distributions do not represent their true underlying population statistics. To combat this, we simulated mock remnant populations for assumed (power law) models of their jet powers and lifetimes, and filtered by the sample selection criteria (\cref{sec:mock_remnant_populations}). Mock catalogues of remnant populations were generated for a range of jet power and lifetime functions and compared to the observed jet powers and active age distributions. We constrained the optimal pairing of jet kinetic power and lifetime functions and compared our results to the theoretical expectations for jet triggering and fuelling mechanisms (\cref{sec:results}). We summarise our findings as follows:
\setitemize{topsep=3pt,parsep=0pt,itemsep=0pt,leftmargin=0pt,itemindent=1.5\parindent}
\begin{itemize}    
    \item We constrain the jet kinetic power and lifetime functions as $p(Q)\propto~Q^{-1.49\pm0.07}$ and $p(t_{\mathrm{on}})\propto~t_{\mathrm{on}}^{-0.97\pm0.12}$, respectively. 

    \item {Our jet lifetime function aligns with the findings of \citet{2019A&A...622A..12H} and \citet{2020MNRAS.496.1706S}, however, our approach allows us to impose more precise constraints on the range of plausible lifetime functions ruling out their $p(t_{\mathrm{on}})\propto~t_{\mathrm{on}}^{-0.5}$ model.}
    
    \item {These lifetime functions are consistent with feedback-regulated accretion simulations, which predict a power-law spectrum in black hole accretion rates equivalent to a $p(t_{\mathrm{on}})\propto t_{\mathrm{on}}^{-1}$ lifetime function.}

    \item {The jet kinetic powers constrained for the majority of our remnant sample (in their active state) are consistent with either a super-Eddington slim disk accretion flow or atypically massive host black holes ($M_\bullet > 10^9$\,M$_\odot$).}

    \item {Our steeper jet power function, compared to \citet{2020MNRAS.496.1706S}, could plausibly result from a different jet triggering mechanism for the order of magnitude more powerful sources in our sample; e.g., cold-mode versus feedback-regulated accretion. We speculate that the true jet power function may be better described by (e.g.) a broken power-law model.}
    

\end{itemize}

Despite the focus of this work being on remnant radio galaxies, our methodology is applicable to active radio galaxy populations (\cref{sec:Application to active radio galaxy populations}). Considering these objects dominate flux and volume limited samples, their statistical significance is an attractive property that should be exploited in similar analyses. 
In the advent of new generation radio surveys such as LoTSS, the {Evolutionary Map of the Universe} \citep[EMU;][]{2011JApA...32..599N}, and eventually those produced by the SKA, constraints on the global energetics of radio-loud AGNs will dramatically improve. Not only will these surveys provide large, statistically-significant samples of radio galaxies, their sensitivity and angular resolution will allow for much lessened selection biases. This will allow, for example, the testing of higher-order models (e.g., double power-law jet kinetic power models), and investigating whether the correlations exist between the jet kinetic powers and lifetimes, or with properties of the host galaxy. Furthermore, with the inclusion of higher-quality optical/infrared \citep[e.g., the LOFAR-WEAVE survey;][]{2016sf2a.conf..271S} and X-ray \citep[e.g., eROSITA;][]{2021A&A...647A...1P} data, our methodology can be extended to explore the environmental impact on AGN fuelling.

\section*{Acknowledgements}

{We thank the anonymous referee for a constructive and prompt report, which helped improve the manuscript.} BQ acknowledges a Doctoral Scholarship and an Australian Government Research Training Programme scholarship administered through Curtin University of Western Australia. NHW is supported
by an Australian Research Council Future Fellowship (project
number FT190100231) funded by the Australian Government. This work was supported by resources provided by the Pawsey Supercomputing Research Centre (\url{https://ror.org/04f2f0537}) with funding from the Australian Government and the Government of Western Australia.

\section*{Data Availability}

The authors confirm that the data supporting the findings of this study are available in cited external data archives or presented in the article. Processed data products underlying this article will be shared on reasonable request to the authors. The relevant code is publicly available as disclosed when referenced in the paper.



\bibliographystyle{mnras}
\bibliography{agnlifecycles} 




\FloatBarrier\appendix
\onecolumn
\begin{figure}
\section{Summary of remnant lobe candidates}
\label{sec:appendix}
The observed and derived properties of the 79 candidates remnant lobes studied in this work are summarised in \cref{tab:table1,tab:table2}.
\end{figure}
\begin{table*}
\begin{tabular}{lcccccccc} 
\hline
Name & Redshift & $S_\text{154}$ & LLS & Jet power & Active age & Halo mass & Remnant ratio & Core \\
 &  & (Jy) & (arcmin) & ($\log$\,W) & ($\log$\,yrs) & ($\log$\,M$_\odot$) & &  \\
\hline
J001748-222305 & 0.1081 & $2.69\pm0.05$ & 8.63 & $38.6\pm0.07$ & $8.2\pm0.03$ & $13.0\pm0.06$ & $0.10\pm0.04$ & no \\
\rowcolor{lightgray!25!white}J003406-663935 & 0.1075 & $0.82\pm0.03$ & 15.86 & $38.5\pm0.08$ & $8.3\pm0.03$ & $12.0\pm0.08$ & $0.03\pm0.01^\dagger$ & yes \\
J003941-130106 & 0.1075 & $1.02\pm0.05$ & 12.5 & $38.8\pm0.02$ & $8.5\pm0.01$ & $13.6\pm0.03$ & $0.05\pm0.01$ & yes \\
J005109-202802 & 0.0856 & $1.09\pm0.03$ & 8.7 & $38.2\pm0.08$ & $8.2\pm0.04$ & $13.0\pm0.05$ & $0.28\pm0.03$ & yes \\
\rowcolor{lightgray!25!white} J005522-485644 & 0.0646 & $0.79\pm0.04$ & 11.9 & $38.3\pm0.03$ & $8.2\pm0.02$ & $12.6\pm0.06$ & $0.00\pm0.01^\dagger$ & yes \\
J011429+050820 & 0.2046 & $0.92\pm0.04$ & 5.96 & $39.2\pm0.04$ & $8.1\pm0.01$ & $13.0\pm0.03$ & $0.03\pm0.01$ & no \\
J011621-472251 & 0.146 & $13.69\pm0.17$ & 12.27 & $40.0\pm0.02$ & $7.9\pm0.03$ & $14.0\pm0.04$ & $0.55\pm0.02$ & no \\
J012323+014513 & 0.0318 & $1.22\pm0.07$ & 8.5 & $37.3\pm0.05$ & $8.0\pm0.04$ & $12.8\pm0.06$ & $0.45\pm0.02$ & no \\
J012932-643342 & 0.1343 & $1.21\pm0.02$ & 6.18 & $38.1\pm0.02$ & $8.3\pm0.01$ & $12.6\pm0.04$ & $0.00\pm0.01^\dagger$ & no \\
J014343-543058 & 0.1791* & $0.69\pm0.03$ & 7.26 & $38.7\pm0.02$ & $8.3\pm0.01$ & $12.6\pm0.03$ & $0.08\pm0.01$ & yes \\
J022151+205426 & 0.162 & $0.57\pm0.05$ & 7.36 & $39.2\pm0.03$ & $7.9\pm0.02$ & $12.8\pm0.06$ & $0.28\pm0.03$ & no$^\text{v}$ \\
\rowcolor{lightgray!25!white} J023431-412907 & 0.0705 & $1.28\pm0.03$ & 8.15 & $38.1\pm0.08$ & $8.3\pm0.03$ & $12.8\pm0.12$ & $0.00\pm0.01^\dagger$ & yes \\
J030957+191251 & 0.0341 & $1.22\pm0.06$ & 6.8 & $37.3\pm0.18$ & $8.1\pm0.12$ & $13.4\pm0.09$ & $0.40\pm0.11$ & yes \\
J033137-771519 & 0.1456 & $1.99\pm0.08$ & 19.11 & $39.3\pm0.07$ & $8.4\pm0.02$ & $12.8\pm0.08$ & $0.03\pm0.01$ & no \\
J035352-603010 & 0.1496* & $0.81\pm0.03$ & 10.3 & $39.1\pm0.01$ & $8.1\pm0.01$ & $12.6\pm0.03$ & $0.05\pm0.01$ & no \\
J040847-624750 & 0.0178 & $3.43\pm0.11$ & 40.37 & $38.6\pm0.02$ & $7.6\pm0.03$ & $12.0\pm0.04$ & $0.60\pm0.01$ & no \\
J043557-233308 & 0.07 & $0.99\pm0.04$ & 7 & $38.1\pm0.08$ & $7.9\pm0.06$ & $12.8\pm0.05$ & $0.50\pm0.04$ & yes \\
J043928-233040 & 0.0746 & $1.17\pm0.04$ & 9.84 & $38.6\pm0.02$ & $8.1\pm0.01$ & $12.4\pm0.04$ & $0.08\pm0.01$ & no$^\text{v}$ \\
J044111+251843 & 0.0548 & $1.98\pm0.1$ & 5.63 & $38.3\pm0.01$ & $8.3\pm0.01$ & $14.4\pm0.01$ & $0.23\pm0.01$ & no$^\text{r}$ \\
J045618-260123 & 0.1401* & $1.17\pm0.03$ & 4.81 & $38.2\pm0.08$ & $8.2\pm0.05$ & $13.4\pm0.08$ & $0.28\pm0.03$ & no \\
J050536-284554 & 0.0381* & $13.51\pm0.16$ & 35.73 & $39.0\pm0.07$ & $8.1\pm0.03$ & $12.8\pm0.05$ & $0.33\pm0.02$ & yes \\
\rowcolor{lightgray!25!white} J051547-805932 & 0.1049* & $1.3\pm0.07$ & 7.8 & $38.5\pm0.09$ & $8.3\pm0.04$ & $13.2\pm0.06$ & $0.05\pm0.1^\dagger$ & yes \\
J054715+010856 & 0.135* & $2.15\pm0.06$ & 8.12 & $38.9\pm0.08$ & $8.2\pm0.02$ & $13.0\pm0.09$ & $0.08\pm0.04$ & yes \\
J064922-330310 & 0.0452* & $1.48\pm0.03$ & 4.63 & $38.0\pm0.07$ & $7.6\pm0.04$ & $12.0\pm0.1$ & $0.05\pm0.01$ & no \\
J065815-382914 & 0.1796* & $1.1\pm0.03$ & 6.02 & $38.4\pm0.08$ & $8.5\pm0.03$ & $13.8\pm0.06$ & $0.10\pm0.03$ & no \\
J072055-440906 & 0.1668* & $2.81\pm0.05$ & 10.48 & $39.6\pm0.04$ & $8.2\pm0.02$ & $13.6\pm0.04$ & $0.13\pm0.02$ & yes \\
\rowcolor{lightgray!25!white} J073037+181702 & 0.1064* & $0.79\pm0.07$ & 9.26 & $38.2\pm0.09$ & $8.4\pm0.04$ & $13.0\pm0.07$ & $0.05\pm0.03^\dagger$ & yes \\
\rowcolor{lightgray!25!white} J073804-442000 & 0.1252* & $1.09\pm0.04$ & 4.58 & $38.0\pm0.04$ & $8.2\pm0.01$ & $12.6\pm0.08$ & $0.00\pm0.01^\dagger$ & yes \\
J074634-570154 & 0.13 & $0.59\pm0.04$ & 7.53 & $38.4\pm0.08$ & $8.2\pm0.03$ & $12.8\pm0.05$ & $0.33\pm0.02$ & yes \\
J075931+082549 & 0.1319 & $0.64\pm0.03$ & 4.47 & $38.3\pm0.04$ & $8.3\pm0.02$ & $13.6\pm0.04$ & $0.05\pm0.01$ & no \\
J083642-110451 & 0.0842* & $1.32\pm0.03$ & 4.66 & $38.4\pm0.01$ & $8.0\pm0.01$ & $12.8\pm0.03$ & $0.03\pm0.01$ & no \\
J084016+175639 & 0.1165 & $3.1\pm0.05$ & 6.15 & $39.3\pm0.07$ & $7.6\pm0.06$ & $13.4\pm0.07$ & $0.70\pm0.08$ & no \\
J092343+242232 & 0.0345 & $1.04\pm0.07$ & 19 & $38.2\pm0.07$ & $8.0\pm0.03$ & $12.2\pm0.06$ & $0.15\pm0.02$ & yes \\
J092739+295704 & 0.0267 & $4.87\pm0.13$ & 12.45 & $37.5\pm0.03$ & $8.2\pm0.01$ & $13.2\pm0.05$ & $0.33\pm0.01$ & no \\
J100029-075448 & 0.0336 & $0.9\pm0.03$ & 4.8 & $36.5\pm0.01$ & $8.3\pm0.01$ & $13.2\pm0.02$ & $0.15\pm0.01$ & no \\
J102130+051928 & 0.1562 & $1.14\pm0.06$ & 15.49 & $39.2\pm0.02$ & $8.5\pm0.01$ & $13.2\pm0.04$ & $0.03\pm0.01$ & yes \\
J102632-395639 & 0.0094 & $1.13\pm0.03$ & 7.74 & $36.8\pm0.02$ & $7.6\pm0.01$ & $13.4\pm0.01$ & $0.38\pm0.01$ & no \\
J103149+225012 & 0.1121 & $1.19\pm0.05$ & 5.28 & $38.2\pm0.06$ & $8.1\pm0.04$ & $13.2\pm0.07$ & $0.38\pm0.03$ & no$^\text{v}$ \\
J103214+275517 & 0.0852 & $1.68\pm0.08$ & 13.88 & $39.2\pm0.05$ & $7.9\pm0.03$ & $12.8\pm0.05$ & $0.35\pm0.02$ & yes \\
J103406+184047 & 0.1366 & $3.17\pm0.08$ & 11.21 & $38.9\pm0.06$ & $8.3\pm0.02$ & $13.0\pm0.05$ & $0.28\pm0.02$ & yes \\
 \hline
\end{tabular}
\caption{Summary of observed and derived properties for the remnant lobe candidates (first 40 of 79 in this table). The columns tabulate the J2000 source name, spectroscopic or photometric redshift, 154\,MHz flux density, largest linear size (LLS), jet power $Q$, active age $t_\text{on}$, halo mass $H$, remnant ratio $R\text{rem}$, and the presence (or not) of a radio core. The remnant candidates highlighted in grey are not confidently classified as remnant lobes either through the RAiSE parameter inversion (2$\sigma$ level) or the `absent radio core' metric; their remnant status is determined based on a combination of the other metrics in \cref{sec:remnant_classification}. \\\textit{Annotations:} * host galaxies with only photometric redshifts; $^\dagger$ remnant ratios statistically consistent with an active lobe (i.e., $R_\text{rem} > 0$ at less than the 2$\sigma$ level); $^\text{r}$ radio sources with jet structure in VLASS that is indicative of a restarted source; $^\text{v}$ faint core detection in VLASS but not RACS-low.}
\label{tab:table1}
\end{table*}

\FloatBarrier\clearpage
\begin{table*}
\begin{tabular}{lcccccccc} 
\hline
Name & Redshift & $S_\text{154}$ & LLS & Jet power & Active age & Halo mass & Remnant ratio & Core \\
 &  & (Jy) & (arcmin) & ($\log$\,W) & ($\log$\,yrs) & ($\log$\,M$_\odot$) & &  \\
\hline
J104844+110808 & 0.157 & $2.02\pm0.13$ & 6.6 & $39.5\pm0.03$ & $7.8\pm0.02$ & $13.6\pm0.04$ & $0.55\pm0.01$ & yes \\
J104945-130802 & 0.15* & $0.83\pm0.04$ & 7.82 & $38.7\pm0.1$ & $8.3\pm0.03$ & $13.0\pm0.07$ & $0.05\pm0.02$ & yes \\
J105732-332445 & 0.0982 & $0.98\pm0.03$ & 6.69 & $39.2\pm0.04$ & $7.7\pm0.02$ & $13.6\pm0.04$ & $0.63\pm0.01$ & no \\
J105915+051730 & 0.0354 & $0.87\pm0.07$ & 14.14 & $37.6\pm0.09$ & $8.3\pm0.04$ & $12.8\pm0.05$ & $0.25\pm0.02$ & yes \\
J105921-170935 & 0.1027 & $0.88\pm0.04$ & 12.2 & $38.5\pm0.08$ & $8.5\pm0.03$ & $13.0\pm0.09$ & $0.08\pm0.01$ & yes \\
J111515-175014 & 0.0677 & $1.19\pm0.04$ & 10.4 & $38.4\pm0.08$ & $7.9\pm0.03$ & $12.6\pm0.06$ & $0.48\pm0.03$ & no \\
J112227-401342 & 0.1809* & $1.45\pm0.03$ & 6.48 & $39.2\pm0.06$ & $8.0\pm0.03$ & $12.8\pm0.07$ & $0.00\pm0.04^\dagger$ & no \\
J112539+082030 & 0.0702 & $0.67\pm0.04$ & 4.1 & $38.5\pm0.02$ & $6.7\pm0.2$ & $11.8\pm0.03$ & $0.70\pm0.01$ & no \\
J113023-345559 & 0.0545* & $0.83\pm0.03$ & 6.02 & $37.7\pm0.09$ & $7.9\pm0.06$ & $13.0\pm0.06$ & $0.53\pm0.05$ & no \\
J113727-005108 & 0.0464 & $1.96\pm0.05$ & 6.1 & $38.5\pm0.15$ & $7.3\pm0.11$ & $12.6\pm0.1$ & $0.68\pm0.08$ & no \\
J115237-303300 & 0.1656* & $3.41\pm0.03$ & 4.81 & $39.8\pm0.02$ & $7.6\pm0.01$ & $14.4\pm0.04$ & $0.60\pm0.01$ & no \\
\rowcolor{lightgray!25!white} J120344+234247 & 0.1767 & $1.95\pm0.07$ & 4.88 & $38.8\pm0.08$ & $8.3\pm0.02$ & $13.6\pm0.05$ & $0.00\pm0.01^\dagger$ & no$^\text{v}$ \\
J121045-435436 & 0.0693 & $0.55\pm0.05$ & 11.17 & $38.6\pm0.19$ & $7.9\pm0.13$ & $12.6\pm0.08$ & $0.43\pm0.16$ & yes \\
J124820-214751 & 0.0419* & $3.13\pm0.06$ & 5.35 & $38.3\pm0.03$ & $7.3\pm0.02$ & $12.8\pm0.04$ & $0.70\pm0.01$ & no \\
J130557-123551 & 0.1639 & $1.27\pm0.06$ & 10.01 & $38.7\pm0.05$ & $8.4\pm0.02$ & $12.8\pm0.03$ & $0.05\pm0.01$ & no \\
J133240-271014 & 0.0399 & $0.55\pm0.06$ & 11.78 & $38.2\pm0.01$ & $8.5\pm0.01$ & $14.4\pm0.01$ & $0.18\pm0.01$ & no \\
J133739-132919 & 0.1448* & $1.21\pm0.05$ & 9.36 & $39.1\pm0.09$ & $8.1\pm0.04$ & $13.2\pm0.07$ & $0.35\pm0.02$ & no \\
J134715-242219 & 0.0199 & $3.58\pm0.05$ & 6.85 & $37.7\pm0.02$ & $7.4\pm0.01$ & $13.0\pm0.05$ & $0.70\pm0.01$ & no \\
J135120-170853 & 0.1755* & $0.57\pm0.03$ & 4.66 & $38.2\pm0.06$ & $8.5\pm0.04$ & $14.0\pm0.05$ & $0.03\pm0.02^\dagger$ & no \\
J143650-161317 & 0.1445 & $1.03\pm0.06$ & 12.9 & $39.0\pm0.09$ & $8.2\pm0.05$ & $12.8\pm0.07$ & $0.30\pm0.03$ & yes \\
J150400-281301 & 0.0858* & $0.45\pm0.04$ & 11.05 & $38.4\pm0.04$ & $8.1\pm0.01$ & $12.0\pm0.04$ & $0.03\pm0.01$ & no$^\text{v}$ \\
J152808+054502 & 0.0411 & $1.25\pm0.07$ & 14.8 & $38.6\pm0.04$ & $7.7\pm0.02$ & $12.6\pm0.04$ & $0.63\pm0.02$ & no$^\text{r}$ \\
J153853+170202 & 0.0296 & $1.23\pm0.1$ & 13.2 & $38.0\pm0.11$ & $7.8\pm0.07$ & $12.8\pm0.07$ & $0.63\pm0.05$ & yes \\
J155903-213902 & 0.077* & $12.44\pm0.06$ & 7.97 & $38.7\pm0.05$ & $8.1\pm0.03$ & $13.6\pm0.05$ & $0.38\pm0.02$ & no \\
J160312-131949 & 0.0801 & $0.95\pm0.05$ & 4.7 & $38.5\pm0.03$ & $7.9\pm0.01$ & $13.6\pm0.02$ & $0.40\pm0.01$ & no$^\text{v}$ \\
J160428-020718 & 0.0304 & $0.54\pm0.06$ & 10.24 & $37.1\pm0.02$ & $8.5\pm0.01$ & $13.6\pm0.03$ & $0.03\pm0.01$ & no \\
J165142-023728 & 0.022 & $2.1\pm0.11$ & 8.48 & $37.3\pm0.02$ & $7.9\pm0.01$ & $13.2\pm0.03$ & $0.50\pm0.01$ & no \\
J165904-030100 & 0.146 & $0.72\pm0.09$ & 9.9 & $38.8\pm0.03$ & $8.5\pm0.01$ & $13.6\pm0.05$ & $0.08\pm0.01$ & no \\
J183119+143802 & 0.1906* & $2.4\pm0.17$ & 15.54 & $39.5\pm0.02$ & $8.5\pm0.01$ & $13.4\pm0.04$ & $0.05\pm0.01$ & yes \\
J191057-704900 & 0.2152* & $1.27\pm0.08$ & 7.85 & $39.6\pm0.02$ & $7.9\pm0.02$ & $13.4\pm0.04$ & $0.43\pm0.02$ & yes \\
J191420-543356 & 0.0171 & $0.65\pm0.06$ & 5.6 & $37.1\pm0.1$ & $7.3\pm0.07$ & $12.2\pm0.07$ & $0.68\pm0.12$ & no \\
J192918-274406 & 0.104 & $1.2\pm0.06$ & 4.4 & $37.7\pm0.13$ & $8.4\pm0.11$ & $13.8\pm0.12$ & $0.10\pm0.01$ & no \\
J195822-373925 & 0.0946* & $1.79\pm0.07$ & 13.05 & $38.5\pm0.07$ & $8.5\pm0.03$ & $13.0\pm0.05$ & $0.05\pm0.01$ & yes \\
J203854-200519 & 0.1284* & $1.04\pm0.04$ & 9.82 & $38.7\pm0.09$ & $8.2\pm0.04$ & $12.6\pm0.08$ & $0.08\pm0.08^\dagger$ & no \\
J210337-681151 & 0.0406 & $0.95\pm0.06$ & 13.29 & $38.3\pm0.12$ & $7.8\pm0.07$ & $12.6\pm0.07$ & $0.63\pm0.05$ & yes \\
J215134-475920 & 0.0629 & $1.04\pm0.02$ & 7.81 & $37.7\pm0.02$ & $8.2\pm0.01$ & $13.0\pm0.03$ & $0.33\pm0.01$ & no \\
\rowcolor{lightgray!25!white} J222350-020626 & 0.0559 & $30.28\pm0.1$ & 8.98 & $38.8\pm0.06$ & $7.5\pm0.13$ & $12.2\pm0.09$ & $0.18\pm0.1^\dagger$ & no$^\text{r}$ \\
J230949-195544 & 0.1087* & $0.54\pm0.02$ & 6.3 & $38.0\pm0.11$ & $8.4\pm0.07$ & $13.0\pm0.11$ & $0.10\pm0.01$ & no \\
J234531-045135 & 0.0756 & $0.83\pm0.06$ & 19.2 & $38.7\pm0.04$ & $8.5\pm0.02$ & $13.2\pm0.06$ & $0.05\pm0.01$ & yes \\
 \hline
\end{tabular}
\caption{Same as \cref{tab:table1} but for the remaining 39 of 79 remnant candidates.}
\label{tab:table2}
\end{table*}


%
%
%
%



\bsp 
\label{lastpage}
\end{document}